\documentclass[aps,prc,nofootinbib,showkeys,showpacs,superscriptaddress,preprint,onecolumn,11pt]{revtex4-2}

\usepackage{graphicx}
\usepackage{dcolumn}
\usepackage{bm}
\usepackage{amsmath,amsfonts,amsthm,amssymb}
\usepackage{hyperref} 
\usepackage{float}
\usepackage{lipsum}
\usepackage{physics}
\usepackage{nicefrac}
\usepackage{yfonts}
\usepackage[utf8]{inputenc}
\usepackage{cancel}

\usepackage[caption=false,labelformat=simple]{subfig}

\usepackage[dvipsnames]{xcolor}

\usepackage[sort&compress]{natbib}
\hypersetup{
colorlinks = true,
linkcolor = blue,
anchorcolor = blue,
citecolor = blue,
filecolor = blue,
urlcolor = blue
}
\usepackage[normalem]{ulem}

\newcommand{\bea}{\begin{eqnarray}}
\newcommand{\eea}{\end{eqnarray}}
\newcommand{\bel}[1]{\begin{eqnarray}\label{#1}}
\newcommand{\eel}{\end{eqnarray}}

\def\LB{\left(}
\def\RB{\right)}
\def\LSB{\left[}
\def\RSB{\right]}
\def\LAB{\langle}
\def\RAB{\rangle}

\newcommand{\nn}{\nonumber}

\newcommand{\EQ}[1]{Eq.~(\ref{#1})}
\newcommand{\EQn}[1]{(\ref{#1})}

\newcommand{\EQS}[1]{Eqs.~(\ref{#1})}

\newcommand{\EQSTWO}[2]{Eqs.~(\ref{#1})~and~(\ref{#2})}
\newcommand{\EQSTWOn}[2]{(\ref{#1})~and~(\ref{#2})}

\newcommand{\EQSM}[2]{Eqs.~(\ref{#1})--(\ref{#2})}
\newcommand{\EQSMn}[2]{(\ref{#1})--(\ref{#2})}

\newcommand{\SEC}[1]{Sec.~\ref{#1}}

\newcommand{\FIG}[1]{Fig.~\ref{#1}}

\newcommand{\CIT}[1]{Ref.~\citep{#1}} 
\newcommand{\CITn}[1]{\citep{#1}}

\newcommand{\Cv}{{\boldsymbol C}}

\newcommand{\dvv}{{\boldsymbol d}} 
\newcommand{\evv}{{\boldsymbol e}}

\def\barr{\begin{array}}
	\def\earr{\end{array}}
\def\beq{\begin{eqnarray}}
	\def\eeq{\end{eqnarray}}
\def\bfig{\begin{figure}}
	\def\efig{\end{figure}}

\newcommand{\p}{\partial}

\newcommand{\f}[2]{\frac{#1}{#2}}

\def\g5{\gamma_5}

\def\S0iU{{\Sigma}^{0i}}

\def\n0{n_{(0)}}
\def\e0{\varepsilon_{(0)}}
\def\P0{P_{(0)}}

\newcommand{\al}{\alpha}

\newcommand{\na}{\nabla}
\newcommand{\om}{\omega}

\begin{document}

\title{Novel hydrodynamic features in dissipative spin-polarized fluids}
            
\author{Zbigniew Drogosz}
\email{zbigniew.drogosz@alumni.uj.edu.pl}
\affiliation{Institute of Theoretical Physics, Jagiellonian University, Kraków, 30-348, Poland}   
          
\author{Wojciech Florkowski}
\email{wojciech.florkowski@uj.edu.pl}
\affiliation{Institute of Theoretical Physics, Jagiellonian University, Kraków, 30-348, Poland}     

\author{Jakub Witkowski}
\email{j.witkowski@student.uj.edu.pl}
\affiliation{Institute of Theoretical Physics, Jagiellonian University, Kraków, 30-348, Poland} 

\begin{abstract}
We investigate an extension of perfect spin hydrodynamics that includes dissipation
in the spin sector. The presence of dissipation implies that antisymmetric parts of the energy--momentum tensor should be taken into account. Their form follows from the Navier--Stokes or Israel--Stewart theory, which introduces a~priori unknown kinetic coefficients and relaxation times. We find that the inclusion of the antisymmetric parts of the energy--momentum tensor has twofold effects: i) they form source terms for the spin tensor allowing for the transfer between the spin and orbital parts of the total angular momentum (spin-orbit interaction) and ii) they appear in the conservation laws for the energy and linear momentum, indirectly affecting the spin dynamics. Numerical results are obtained for two boost-invariant expansion geometries: one-dimensional longitudinal motion (Bjorken model) and two-dimensional boost-invariant and cylindrically symmetric expansion. They indicate (for various choices of the spin kinetic coefficients and the relaxation times) that the spin polarization tensor approaches thermal vorticity. We also observe that Israel--Stewart dynamics on the way to local equilibrium approaches first the Navier--Stokes regime. In general, we find that the Israel--Stewart formalism naturally helps to consistently combine the conservation laws for linear and angular momentum, while the Navier--Stokes framework itself can lead to a disruption in the structure of the equations through the inclusion of terms involving derivatives of an order higher than one.
\end{abstract}

\maketitle

\section{Introduction}
\label{introduction}

Phenomena connected with spin polarization and alignment attract a lot of attention currently~\CITn{Liang:2004ph, Liang:2004xn, STAR:2017ckg, STAR:2018gyt, STAR:2019erd, ALICE:2019aid}; for a recent experimental review, see~\CITn{Niida:2024ntm}. From the theoretical point of view, the spin polarization effects initiated interest in spin hydrodynamics~\CITn{Becattini:2009wh, Becattini:2011zz, Montenegro:2017rbu, Florkowski:2017ruc, Florkowski:2017dyn, Florkowski:2018ahw, Hattori:2019lfp, Weickgenannt:2019dks, Hattori:2019lfp, Florkowski:2019qdp, Bhadury:2020puc, Montenegro:2020paq, Weickgenannt:2020aaf, Shi:2020htn, Bhadury:2020cop, Fukushima:2020ucl, Li:2020eon, Singh:2020rht, Gallegos:2021bzp, Weickgenannt:2021cuo, She:2021lhe, Hongo:2021ona, Hu:2021pwh, Singh:2022ltu, Weickgenannt:2022zxs, Wagner:2022amr, Dey:2023hft, Weickgenannt:2023nge, Kumar:2023ojl, Wagner:2024fhf, Wagner:2024fry, Dey:2024cwo, She:2024rnx, Huang:2024ffg, Bhadury:2025fil, Dey:2025wqw, Daher:2026psy}. This framework includes spin degrees of freedom into the well-established formalism of relativistic hydrodynamics -- the basic tool to interpret the results of heavy-ion experiments~\CITn{Ollitrault:2007du, Romatschke:2009im, Florkowski:2010zz, Gale:2013da, Jaiswal:2016hex}.

In this work we study an extension of perfect spin hydrodynamics that includes dissipation in the spin sector. The framework of perfect spin hydrodynamics follows our recent formulation summarized in~\CITn{Florkowski:2024cif} and~\CITn{Bhadury:2025dzh}. The inclusion of dissipation is achieved by adding antisymmetric parts to the energy--momentum tensor~\CITn{Hattori:2019lfp, Drogosz:2024gzv}. This is done within the Navier--Stokes (NS)~\CITn{Hattori:2019lfp, Drogosz:2024gzv} or Israel--Stewart (IS)~\CITn{Biswas:2023qsw} approach. In both cases one has to introduce kinetic coefficients whose forms are generally unknown. We assume that they scale with the entropy density and consider different proportionality constants. Similarly, using the IS theory, we consider in addition different values of the spin relaxation times. 

The antisymmetric parts of the energy--momentum tensor have twofold effects. First, they represent source terms for the spin tensor, which allows for the transfer between the spin and orbital parts of the total angular momentum (dissipative spin-orbit interaction). Second, they appear in the conservation laws for the energy and linear momentum, indirectly affecting spin dynamics. In our opinion, such a feedback has not yet been sufficiently analyzed. We find that it is highly restrictive for spin evolution. 

Our numerical results are obtained for two boost-invariant geometries: a one-dimensional longitudinal motion and a two-dimensional boost-invariant and cylindrically symmetric expansion. To distinguish between these two cases, we use the acronyms TH (transversely homogeneous system) and CS (cylindrically symmetric system). The antisymmetric part of the energy--momentum tensor gives a contribution to the entropy production that involves a difference between the spin polarization tensor $\omega$ and the thermal vorticity tensor $\varpi$. As the system approaches local equilibrium, the entropy production rate decreases, and we expect that the spin polarization tensor approaches thermal vorticity. This behavior is confirmed by our numerical results.

In order to concentrate on spin effects mainly, we ignore dissipation connected with heat conduction, shear viscosity, and bulk viscosity. Nevertheless, in the spin sector we find many similarities to the standard dissipative dynamics. For example, on the way to local equilibrium the IS dynamics approaches first the NS regime. On the other hand, it turns out that the IS formalism is particularly convenient for describing dissipative spin dynamics, as it allows for the direct determination of the time derivative of the antisymmetric components of the energy--momentum tensor, which appear in the conservation laws for energy and linear momentum. However, the use of the NS approach alone may lead to an undesired  modification of hydrodynamic equations via introduction of terms involving derivatives of an order higher than one.

\bigskip
Our present work can be interpreted as a continuation of a series of previous studies~\CITn{Florkowski:2019qdp, Florkowski:2021wvk, Wang:2021wqq, Singh:2021man,  Singh:2020rht, Wang:2021ngp,  Drogosz:2024lkx, Drogosz:2026egr, Singh:2026ytd, Drogosz:2026qbo} that offered numerical and analytic solutions to the equations of spin hydrodynamics (such studies analyzed, for example, one dimensional boost-invariant expansion~\CITn{Florkowski:2019qdp, Wang:2021ngp, Singh:2021man, Drogosz:2024lkx, Singh:2026ytd, Drogosz:2026qbo}, two-dimensional Gubser expansion~\CITn{Singh:2020rht, Wang:2021wqq}, one-dimensional non boost-invariant motion~\CITn{Florkowski:2021wvk}, boost-invariant and cylindrically symmetric geometry~\CITn{Drogosz:2026egr}). Our results shed light on the future more realistic applications of dissipative spin hydrodynamics. In particular, they emphasize the advantages of using the IS theory.

\bigskip 
Our paper is organized as follows: In~\SEC{sec:EMTandST} we introduce  expressions for the energy--momentum and spin tensors that include dissipative terms. In~\SEC{sec:Entropy} the entropy production is discussed in the NS and IS approaches. Equations defining the dissipative terms of the energy--momentum tensor are given for these two cases. Boost-invariant and transversely homogeneous systems are discussed in~\SEC{sec:TH}, while boost-invariant and cylindrically symmetric systems are investigated in~\SEC{sec:CS}. We present a summary in~\SEC{sec:summary}, and a collection of useful differential identities is given in Appendix \ref{sec:BICSuseful}.

\bigskip
{\it Notation and conventions}: For the Levi-Civita tensor
$\epsilon^{\mu\nu\alpha\beta}$ we follow the convention $\epsilon^{0123} =-\epsilon_{0123} = +1$. The metric tensor is of the ``mostly minuses'' form $g_{\mu\nu} = \textrm{diag}(+1,-1,-1,-1)$. We use boldface symbols to denote three-vectors, and their scalar products are written as $\bm{a}\cdot\bm{b}$. The particles are always considered on the mass shell with the energy $E_{\bm{p}} = \sqrt{m^2+\bm{p}^2}$. Throughout the text, we make use of natural units, $\hbar = c = k_{\rm B} = 1$.

We also use the standard notation where round (squared) brackets denote symmetrization (antisymmetrization) of the Lorentz indices:  $A^{(\mu\nu)}= \frac{1}{2}(A^{\mu\nu}+A^{\nu\mu})$ and $A^{[\mu\nu]}=  \frac{1}{2} (A^{\mu\nu}-A^{\nu\mu})$. Moreover, we introduce an orthogonal projection $T^{\alpha \beta \LAB \gamma \RAB \delta ...} = \Delta^\gamma_\rho T^{\alpha \beta \rho \delta ...}$, where the operator $\Delta^\gamma_\rho = g^\gamma_\rho - U^{\gamma} U_{\rho}$ projects on the plane transverse to the flow vector $U$. If the symbols $\LAB \RAB$ are used to the right of the differential symbol, the contractions with $\Delta$ should be done after the derivative is calculated first. The transverse gradient $\nabla^\mu$ is defined by the contraction $\Delta^{\mu \nu} \p_\nu$.

\section{Energy--momentum and spin tensors}
\label{sec:EMTandST}

\subsection{Asymmetric energy--momentum tensor}

In this work, we consider a system whose energy--momentum tensor has a symmetric and an antisymmetric part, $T^{\mu\nu} = T^{\mu\nu}_{\rm eq} + \delta T^{\mu\nu}_a$. The symmetric part, $T^{\mu\nu}_{\rm eq}$, describes local equilibrium, while the antisymmetric part, $\delta T^{\mu\nu}_a$, is dissipative but includes only the spin dissipation effects. Since we intend to isolate and study the consequences of having an antisymmetric part of the energy--momentum tensor, we neglect standard viscous corrections to $T^{\mu\nu}_{\rm eq}$ that are symmetric. 

The tensor $T^{\mu\nu}_{\rm eq}$ has the well known perfect-fluid form
\begin{eqnarray}
T^{\mu\nu}_{\rm eq} = \LB \varepsilon + P \RB U^\mu U^\nu - P g^{\mu\nu},   
\label{eq:perfect}
\end{eqnarray}
where $\varepsilon$ is the energy density, $P$ is the equilibrium pressure, and $U^\mu$ is the hydrodynamic flow vector (with the normalization $U^\mu U_\mu = 1$). The energy density and pressure are connected through the equation of state. They can be treated as functions of temperature $T$ and baryon chemical potential $\mu$. In numerical calculations, we use the expressions describing a relativistic gas of massive particles (for their explicit forms see, for example, our previous work~\CITn{Drogosz:2026egr}). Following earlier studies, we neglect dependence of $T^{\mu\nu}_{\rm eq}$ on the components of the spin polarization tensor $\omega^{\mu\nu}$, as such dependence appears only as the quadratic and higher order corrections in the components of $\omega^{\mu\nu}$.

The antisymmetric part of $T^{\mu\nu}$ can be in general written in the form~\CITn{Hattori:2019lfp, Drogosz:2024gzv}
\begin{eqnarray}
\delta T^{\mu\nu}_a = d^\mu U^\nu - d^\nu U^\mu + e^{\mu\nu}.  \label{eq:deltaTa}  
\end{eqnarray}
The vector $d^\mu$ is orthogonal to the flow vector, $d \cdot U = 0$, while the tensor $e^{\mu\nu}$ is antisymmetric and orthogonal to $U$ in two indices, $ e^{\mu\nu} U_\mu = e^{\mu\nu} U_\nu = 0$. There are three independent components in each of $d^\mu$ and $e^{\mu\nu}$ -- together, they represent the six independent components of $\delta T^{\mu\nu}_a$. The properties of $e^{\mu\nu}$ allow us to represent it as the contraction of the Levi-Civita tensor with the hydrodynamic flow vector $U$ and a vector $e$ that is orthogonal to $U$, namely
\begin{equation}
 e^{\mu\nu} = \varepsilon^{\mu \nu \alpha \beta} U_\alpha e_\beta, \quad e \cdot U = 0.  
 \label{eq:emunu}
\end{equation}
In analogy to magnetohydrodynamics, we may think of the vectors $d$ and $e$ as the electriclike and magneticlike contributions to $\delta T^{\mu\nu}_a$.

The energy and linear momentum conservation is expressed by the equation
\begin{eqnarray}
\p_\mu   T^{\mu\nu} =  \p_\mu   T^{\mu\nu}_{\rm eq}   + \p_\mu  \delta T^{\mu\nu}_a = 0.
\label{eq:enmomcon}
\end{eqnarray}
Below, we consider two expansion geometries connected with two different sets of basis vectors ${\cal X}^\mu_{(i)} \, (i=0, 1,2,3)$. By projecting \EQ{eq:enmomcon} along ${\cal X}^\mu_{(i)}$, we obtain four equations
\begin{eqnarray}
{\cal X}_\nu^{(i)} \, \p_\mu   T^{\mu\nu} = 0.
\label{eq:enmomconX}
\end{eqnarray}
As the first basis vector, we always choose the flow vector, ${\cal X}_\nu^{(0)} = U_\nu$.

\subsection{Spin tensor}

Similarly to the energy--momentum tensor, we decompose the spin tensor into the equilibrium part and a correction, $S^{\lambda, \mu \nu} = S^{\lambda, \mu \nu}_{\rm eq} + \delta S^{\lambda, \mu \nu}$. The equilibrium part has the form~\CITn{Florkowski:2017ruc}
\begin{equation}
S^{\lambda, \mu \nu}_{\rm eq} = U^\lambda\big[A(k^\mu U^\nu - k^\nu U^\mu) + A_1 t^{\mu\nu}\big] + \frac{A}{2}\big(t^{\lambda\mu} U^\nu - t^{\lambda \nu} U^\mu + \Delta^{\lambda \mu} k^\nu - \Delta^{\lambda \nu}k^\mu \big),
\label{eq:Sglw}
\end{equation}
where $A$ and $A_1$ are functions of $T$ and $\mu$ (explicitly defined, for example, in~\CITn{Drogosz:2026egr}), whereas the vector $k$ and the tensor $t^{\lambda\mu} $ define the spin polarization tensor
\bea
\omega_{\mu \nu}=k_{\mu} U_{\nu} -k_{\nu} U_{\mu}+t_{\mu\nu} =
k_{\mu} U_{\nu} -k_{\nu} U_{\mu}+\epsilon_{\mu \nu \alpha \beta} U^{\alpha} \omega^{\beta}.
\label{eq:o1}
\eea
The vectors $k$ and $\omega$ satisfy the orthogonality conditions, $k \cdot U = 0$ and $\omega \cdot U = 0$.

The correction to the equilibrium part of the spin tensor has the form~\CITn{Drogosz:2024gzv}
\bea
\delta S^{\lambda, \mu \nu} = 
\delta i^{\lambda \mu} U^\nu -
\delta i^{\lambda \nu} U^\mu +
\delta j^{\lambda \mu \nu},
\label{eq:didj}
\eea
where the tensors $\delta i^{\lambda \mu} $ and $\delta j^{\lambda \mu \nu}$ satisfy the following relations: $\delta i^{\lambda \mu} U_\lambda = \delta i^{\lambda \mu} U_\mu = 0$, $\delta j^{\lambda \mu \nu} = -\delta j^{\lambda \nu \mu}$, $\delta j^{\lambda \mu \nu} U_\lambda = \delta j^{\lambda \mu \nu} U_\mu = \delta j^{\lambda \mu \nu} U_\nu = 0$ (consequently, there are nine independent components in $\delta i^{\lambda \mu}$ and also nine independent components in $\delta j^{\lambda \mu \nu}$).

Since the energy--momentum tensor has an antisymmetric part, it acts as the source term for the spin tensor, namely, the conservation of the total angular momentum can be written as
\begin{eqnarray}
\p_\lambda  S^{\lambda, \mu\nu} =  \p_\lambda   S^{\lambda, \mu\nu}_{\rm eq}   + \p_\lambda  \delta S^{\lambda, \mu\nu} = \delta T^{\nu\mu}_a - \delta T^{\mu\nu}_a .
\label{eq:spincon}
\end{eqnarray}
%

\section{Entropy production}
\label{sec:Entropy}

\subsection{Navier--Stokes limit} 

In the Navier--Stokes limit, the entropy production is given by the expression~\CITn{Hattori:2019lfp, Drogosz:2024gzv}
\bel{eq:divS}
\p_\mu S^\mu_{\rm NS} =  
- \delta N^\mu \p_\mu \xi
+ \delta T^{\mu \lambda}_s \p_\mu \beta_\lambda 
+ \delta T^{\mu \lambda}_a \LB \p_\mu \beta_\lambda 
- \omega_{\lambda\mu} \RB
-\f{1}{2} \delta S^{\mu, \alpha \beta} \p_\mu \omega_{\alpha\beta} ,
\eel
where $\delta N^\mu$ is the nonequilibrium correction to the baryon current, $\delta T^{\mu \lambda}_s$ ($\delta T^{\mu \lambda}_a$) is the nonequilibrium symmetric (antisymmetric) correction to the energy--momentum tensor, and $\delta S^{\mu, \alpha \beta}$ is the nonequilibrium correction to the spin tensor. 

Since we are only interested in the spin effects, we set $\delta N^\mu = 0$ and $\delta T^{\mu \lambda}_s = 0$. We also assume here that $\delta S^{\mu, \alpha \beta} = 0$. In this case, the formula for the entropy production reduces to
\bel{eq:divSNS1}
\p_\mu S^\mu_{\rm NS} =  
\delta T^{\mu \lambda}_a \LB \p_\mu \beta_\lambda 
- \omega_{\lambda\mu} \RB =
\delta T^{\mu \lambda}_a \LB 
\varpi_{\lambda\mu}
- \omega_{\lambda\mu} \RB,
\eel
where we have introduced the thermal vorticity
\begin{equation}
\varpi_{\mu\nu} = -\f{1}{2} \LB
\p_\mu \beta_\nu - \p_\nu \beta_\mu \RB. 
\label{eq:varpi}
\end{equation}
Using the decomposition \EQn{eq:deltaTa} we find
\bel{eq:divSNS2}
\p_\mu S^\mu_{\rm NS} &=&  d^\mu \LB 2 k_\mu - \beta a_\mu - \beta^2 \nabla_\mu T \RB +  e^{\mu\nu} \LB  \beta \nabla_{[\mu} U_{\nu ]}
+  \epsilon_{\mu\nu\alpha\beta} U^\alpha \omega^\beta \RB,
\eel
where $a^\mu = D U^\mu = U^\lambda \p_\lambda U^\mu$ is the four-acceleration of the fluid element. The positivity of the entropy production is guaranteed by the conditions
\begin{eqnarray}
d^\mu = d^\mu_{\rm NS} &=& \lambda \beta^{-1} 
\LB \beta a^\mu + \beta^2  \nabla^\mu T - 2 k^\mu \RB, 
\label{eq:da} \\
e^{\mu \nu} = e^{\mu \nu}_{\rm NS} &=&  \gamma  \beta^{-1} \LB \beta
\nabla^{[ \mu} U^{\nu ]}
+  \epsilon^{\mu\nu\alpha\beta} U_\alpha \omega_\beta \RB.
\label{eq:ea} 
\end{eqnarray}
The coefficients $\lambda$ and $\gamma$ are the ``new'' (nonnegative) spin kinetic coefficients introduced for the first time in~\CITn{Hattori:2019lfp}. We have rescaled them compared to~\CITn{Hattori:2019lfp} in such a way that their dimension agrees with the dimension of the entropy density $\sigma$ (in numerical calculations we assume proportionality between $\lambda$, $\gamma$ and $\sigma$, and consider different proportionality constants between these variables). Within our assumptions, \EQSTWO{eq:da}{eq:ea} should be used in the dynamic equations
\begin{eqnarray}
\p_\mu  \LB  T^{\mu\nu}_{\rm eq}   + d^\mu U^\nu - d^\nu U^\mu + e^{\mu\nu} \RB &=& 0,
\label{eq:conNSenergy} \\
\p_\lambda   S^{\lambda, \mu\nu}_{\rm eq} + 2 \LB d^\mu U^\nu - d^\nu U^\mu + e^{\mu\nu} \RB &=& 0.
\label{eq:conNSspin}
\end{eqnarray}

\subsection{Israel--Stewart approach}

In the Israel--Stewart approach~\CITn{Israel:1976efz, Israel:1979wp}, the entropy production is defined by the formula
\bel{eq:divSIS}
\p_\mu S^\mu_{\rm IS} =  
\delta T^{\mu \lambda}_a \LB \varpi_{\lambda\mu} 
- \omega_{\lambda\mu} \RB
 + \p_\mu Q^\mu,
\eel
where the four-vector $Q^\mu$ is defined by the expression~\CITn{Biswas:2023qsw}~\footnote{Here we consider only two out of many possible terms given in~\CITn{Biswas:2023qsw}. The role of other terms will be discussed in a separate work.}
\begin{equation}
Q^\mu = a_4 U^\mu d^\lambda d_\lambda + a_5 U^\mu e^{\alpha\beta } e_{\alpha\beta}. 
\end{equation}
In this case, the vector $d$ satisfies the equation
\begin{equation}
 \tau_d\,  {\dot d}^{\, \langle \mu \rangle}_{\rm IS} + d^{\,\mu }_{\rm IS} 
=  - \lambda T \LB 2 k^\mu - \beta a^\mu - \beta^2  \nabla^\mu T+ d^\mu_{\rm IS} \,{U^\lambda \p_\lambda a_4}  + a_4 \Theta d^\mu_{\rm IS} \RB,
\label{eq:IS_d_gen}
\end{equation}
where $\Theta \equiv \partial_\mu U^\mu$ is the expansion scalar, $\tau_d = 2 \lambda T a_4$ is the relaxation time, while the tensor $e^{\mu\nu }$ fulfills the equation
\begin{equation}
\tau_e\,  {\dot e}^{\, \langle [\mu \nu] \rangle}_{\rm IS} + e^{\,\mu\nu }_{\rm IS} 
= \gamma \beta^{-1}  \left( \beta \nabla^{[ \mu} U^{\nu ]}
+  \epsilon^{\mu\nu\alpha\beta} U_\alpha \omega_\beta + e^{\,\mu\nu }_{\rm IS}  \, {U^\lambda \p_\lambda a_5}  
+ a_5 \Theta e^{\,\mu\nu }_{\rm IS} 
\right),
\label{eq:IS_e_gen}
\end{equation}
where $\tau_e = -2 \gamma T a_5$ is another relaxation time. We note that $a_4 > 0$, while $a_5 < 0$.

\section{Boost-invariant and transversely uniform system}
\label{sec:TH}

In our numerical analyses, we first consider the Bjorken expansion~\CITn{Bjorken:1982qr}, i.e., a one-dimensional, boost-invariant expansion of a system that is uniform in the transverse plane. We call this geometry of expansion shortly the TH (transversely homogeneous) model. 

\subsection{Basis vectors and basic symmetries}
\label{sec:bitubasis}

In the TH model, it is convenient to use a basis ${\cal X}^\mu_{(i)}$ whose vectors have a form invariant under boosts along the $z$ axis. A natural choice is~\CITn{Florkowski:2011jg}:
\begin{align}\begin{split}
{\cal X}^\mu_{(0)} = U ^\mu &= \frac{1}{\tau}(t,0,0,z)= ( \cosh\eta,0,0, \sinh\eta), \\
{\cal X}^\mu_{(1)} = X^\mu &= (0,1,0,0), \\
{\cal X}^\mu_{(2)} = Y^\mu &= (0,0,1,0), \\
{\cal X}^\mu_{(3)} = Z^\mu &= \frac{1}{\tau}(z,0,0,t)= ( \sinh\eta,0,0,\cosh\eta).\label{eq:basis}
\end{split}\end{align}
Here $\tau=\sqrt{t^2 - z^2}$ is the longitudinal proper time, and $\eta= \frac{1}{2} \ln \frac{t+z}{t-z}$ is the spacetime rapidity. The four-vector $U^\mu$ is timelike, while the four-vectors $X^\mu, Y^\mu, Z^\mu$ are spacelike. In addition, they are pairwise orthogonal, 
\begin{eqnarray}
U^2 &=& 1, \quad X^2 = Y^2 = Z^2 = -1, \nonumber \\
U \cdot X &=& 0, \quad U \cdot Y = 0, \quad U \cdot Z = 0, \nonumber \\
Z \cdot Y &=& 0, \quad Y \cdot X = 0, \quad X \cdot Z = 0.
\label{eq:normTH}
\end{eqnarray}

The basis vectors \EQn{eq:basis} satisfy also the identities:
\begin{align}\begin{split}
\epsilon_{\alpha\beta\gamma\delta} U^\beta X^\gamma Y^\delta = - Z_\alpha, \\
\epsilon_{\alpha\beta\gamma\delta} U^\beta Y^\gamma Z^\delta = - X_\alpha, \\
\epsilon_{\alpha\beta\gamma\delta} U^\beta Z^\gamma X^\delta = - Y_\alpha. \\
\label{eq:basisLC}
\end{split}\end{align}
Boost-invariant scalar functions (such as thermodynamic quantities)  depend only on the proper time. For the derivatives of scalar functions with respect to $\tau$, we introduce the dot notation
\begin{equation}
D f(\tau) \equiv U^\mu \p_\mu f(\tau) = \frac{\dd f(\tau)}{\dd \tau} \equiv {\dot f}(\tau).
\label{eq:dotnot}
\end{equation}
The projector $\Delta^{\mu \nu} \equiv g^{\mu \nu} - U^{\mu} U^{\nu}$, orthogonal to the flow in both indices, is expressed in the basis~(\ref{eq:basis}) as $\Delta^{\mu \nu} = - X^\mu X^\nu - Y^\mu Y^\nu- Z^\mu Z^\nu$. Using this formula, it is easy to check that the transverse gradient of scalar functions vanishes
\begin{equation}
\nabla^\mu f(\tau) =   \Delta^{\mu \nu} \p_\nu f(\tau) = 0.
\label{eq:nablaf}
\end{equation}

Using the properties listed above, one finds that the following conditions are satisfied in the TH model
\begin{eqnarray}
a^\mu = 0, \quad  \nabla^\mu T = 0, \quad    \nabla^{[ \mu} U^{\nu ]} = 0, \quad \varpi^{\mu\nu} = 0.
\label{eq:THcond}
\end{eqnarray}
Hence, in the TH model the Navier--Stokes expressions \EQn{eq:da} and \EQn{eq:ea} are reduced to simple expressions
\begin{eqnarray}
d^\mu_{\rm NSTH} &=& - 2 k^\mu \lambda \beta^{-1}, 
\label{eq:daTH} \\
e^{\mu \nu}_{\rm NSTH} &=&  \gamma  \beta^{-1} \epsilon^{\mu\nu\alpha\beta} U_\alpha \omega_\beta,
\label{eq:eaTH} 
\end{eqnarray}
with the entropy production given by the formula
\bel{eq:divSNSTH}
\p_\mu S^\mu_{\rm NSTH} &=& -4 \lambda \beta^{-1} k^\mu k_\mu - 2 \gamma \beta^{-1} \omega^\mu \omega_\mu,
\eel 
which is manifestly positive, since $k^\mu k_\mu \leq 0$ and $\omega^\mu \omega_\mu \leq 0$. It is important to notice that on the right-hand sides of \EQSTWO{eq:daTH}{eq:eaTH} no terms with gradients of hydrodynamic variables appear. This is so because the entropy production is proportional to the difference between the spin polarization tensor and the thermal vorticity, and the latter vanishes in the TH model. Consequently, we expect that dissipation in the TH model leads to destruction of the spin polarization, if such is present at the initial stage of the system evolution.

Given the orthogonality conditions, $k \cdot U = 0$ and $\omega \cdot U = 0$, one can introduce the following decompositions~\CITn{Florkowski:2019qdp}:
\begin{align}
k^{\mu} &= C_{kx}X^{\mu} + C_{ky}Y^{\mu} + C_{kz}Z^{\mu} = \left( C_{kz}  \sinh \eta, C_{kx}, C_{ky}, C_{kz} \cosh\eta \right),\\
\omega^{\mu} &= C_{\omega x}X^{\mu} + C_{\omega y}Y^{\mu} + C_{\omega z}Z^{\mu} = \left( C_{\omega z}  \sinh\eta, C_{\omega x}, C_{\omega y}, C_{\omega z} \cosh\eta \right),
\label{eq:CkCo}
\end{align}
where the coefficient functions $C_{k i}$ and $C_{\omega i} \, (i=x,y,z)$ depend on the proper time only. Analogous decompositions hold for the vectors $d$ and $e$:
\begin{align}
d^{\mu} &= d_{x}X^{\mu} + d_{y}Y^{\mu} + d_{z}Z^{\mu} = \left( d_{z}  \sinh \eta, d_{x}, d_{y}, d_{z} \cosh\eta \right),\\
e^{\mu} &= e_{x}X^{\mu} + e_{y}Y^{\mu} + e_{z}Z^{\mu} = \left( e_{z}  \sinh \eta, e_{x}, e_{y}, e_{z} \cosh\eta \right),
\label{eq:dw}
\end{align}
where the coefficient functions $d_i$ and $e_i$ also depend only on the proper time. The properties of the tensor $e^{\mu\nu}$ allow us to represent it also in the form
\begin{equation}
e^{\mu\nu} = -e_x(\tau) \LB Y^\mu Z^\nu - Y^\nu Z^\mu \RB  - e_y(\tau) \LB Z^\mu X^\nu - Z^\nu X^\mu \RB -
e_z(\tau) \LB X^\mu Y^\nu - X^\nu Y^\mu \RB.
\label{eq:emunuXYZ}
\end{equation}
The equivalence of \EQ{eq:emunuXYZ} with \EQ{eq:emunu} can be directly established with the help of the identities~\EQn{eq:basisLC}. Below, we frequently use the three-vector notation
\begin{eqnarray}
\Cv_k = \LB C_{kx}, C_{ky},  C_{kz} \RB,&& \quad
\Cv_\omega = \LB C_{\omega x}, C_{\omega y}, C_{\omega z} \RB, \nn \\
\dvv = \LB d_x, d_y,  d_z \RB,&& \quad
\evv = \LB e_x, e_y,  e_z \RB.
\label{eq:3vecnot}
\end{eqnarray}
From now on, we will refer to the vectors $\dvv$ and $\evv$ as to the {\it electric and magnetic dissipation vectors}, or shortly, as to the electric and magnetic dissipation.

\subsection{Conservation laws}
\label{sec:bituEMcon}

\subsubsection{Baryon number, energy, linear momentum}

Since we consider a system with finite baryon number density $n$, we have to include the baryon number conservation law, $\p_\mu (n U^\mu) = 0$, which for the TH model takes the form
\begin{equation}
\f{\dd n}{\dd\tau} + \f{n}{\tau} = 0. 
\label{eq:BNcon}
\end{equation}
The energy--momentum conservation equation~\EQn{eq:conNSenergy} contracted with ${\cal X}_\nu^{(i)}$ gives additional four equations,
\begin{eqnarray}
 {\dot \varepsilon} +\frac{\varepsilon+P}{\tau} &=& 0, \label{eq:E0TH} \\
 {\dot d_x} + \f{d_x}{\tau} &=& 0, \label{eq:E1TH} \\
 {\dot d_y} + \f{d_y}{\tau} &=& 0, \label{eq:E2TH} \\
 {\dot d_z}  &=& 0.
\label{eq:E3TH}
\end{eqnarray}
A new and striking feature of the equations above, which can be regarded as equations of motion of the fluid, is the presence of the coefficients of the vector $\dvv$ in them. This is a consequence of an asymmetric energy--momentum tensor. Interestingly, the components of the tensor $e^{\mu\nu}$ do not appear in \EQSM{eq:E0TH}{eq:E3TH}, since $\p_\mu e^{\mu\nu} = 0$. This property follows from the structure of~\EQ{eq:emunuXYZ}.

We note that the solutions of \EQ{eq:BNcon} and \EQSM{eq:E0TH}{eq:E3TH} have simple forms
\begin{equation}
n(\tau) = \f{\tau_0}{\tau} n^0, \quad 
d_x(\tau) = \f{\tau_0}{\tau} d_x^{\,0}, \quad 
d_y(\tau) = \f{\tau_0}{\tau} d_y^{\,0}, \quad 
d_z(\tau) = d_z^{\,0},
\label{eq:dsol}
\end{equation}
where $n^0, d_x^{\,0}, d_y^{\,0}$ and $d_z^{\,0}$ are constants, while $\tau_0$ is the initial proper time for the hydrodynamic evolution. With the help of the thermodynamic identities $\dd P = T \dd\sigma + \mu \dd n$ and $\varepsilon + P = T \sigma + \mu n$, where $\sigma$ is the entropy density, one can show that \EQ{eq:E0TH} is reduced to the entropy conservation in the form $\dd\sigma/\dd\tau + \sigma/\tau = 0$, again with the scaling solution $\sigma = \sigma_0 \tau_0/\tau$.

\subsubsection{Angular momentum}

The angular momentum conservation equation \EQn{eq:conNSspin} contracted with pairs of the basis vectors $U_\mu X_\nu$, $U_\mu Y_\nu$, $U_\mu Z_\nu$, $Y_\mu Z_\nu$, $Z_\mu X_\nu$, and $X_\mu Y_\nu$ gives six equations:

\begin{equation}\label{eq:cs}
\begin{bmatrix}
A & 0 & 0 & 0 & 0 & 0 \\
0 & A & 0 & 0 & 0 & 0 \\
0 & 0 & A & 0 & 0 & 0 \\
0 & 0 & 0 & A_1 & 0 & 0 \\
0 & 0 & 0 & 0 & A_1  & 0 \\
0 & 0 & 0 & 0 & 0 & A_1
\end{bmatrix}
\begin{bmatrix}
\Dot{C}_{k x} \\
\Dot{C}_{k y} \\
\Dot{C}_{k z} \\
\Dot{C}_{\om x} \\
\Dot{C}_{\om y} \\
\Dot{C}_{\om z} \end{bmatrix}=\begin{bmatrix}
Q_x & 0 & 0 & 0 & 0 & 0 \\
0 & Q_y & 0 & 0 & 0 & 0 \\
0 & 0 & Q_z & 0 & 0 & 0 \\
0 & 0 & 0 & R_x & 0 & 0 \\
0 & 0 & 0 & 0 & R_y  & 0 \\
0 & 0 & 0 & 0 & 0 & R_z
\end{bmatrix}
\begin{bmatrix}
{C}_{k x} \\
{C}_{k y} \\
{C}_{k z} \\
{C}_{\om x} \\
{C}_{\om y} \\
{C}_{\om z} \end{bmatrix}
- 2
\begin{bmatrix}
{d}_{x} \\
{d}_{y} \\
{d}_{z} \\
e_x \\
e_y \\
e_z \end{bmatrix}. 
\end{equation}

\noindent
Here the functions $A$ and $A_1$ define spin densities. The other functions are defined through the relations
\begin{align}
Q_x = Q_y &= -\LB \dot A + \frac{3A}{2\tau} \RB, \quad &&Q_z = -\LB \dot A + \frac{A}{\tau} \RB,\\
R_x = R_y &= -\LB \dot A_1 + \frac{2 A_1 - A}{2\tau} \RB, \quad 
&&R_z = -\LB \dot A_1 + \frac{A_1}{\tau} \RB.
\end{align}
Equations~\EQn{eq:BNcon}, \EQSMn{eq:E0TH}{eq:E3TH}, and \EQn{eq:cs} represent a  system of eleven equations for fourteen unknown functions of the proper time: $T$, $\mu$, $d_x$, $d_y$, $d_z$, $e_x$, $e_y$, $e_z$, $C_{kx}$, $C_{ky}$, $C_{kz}$, $C_{\omega x}$, $C_{\omega y}$, and $C_{\omega z}$. The main question now is, however, how they can be solved together with the constraints following from the entropy production condition. This issue is discussed in the following section.

\subsection{Entropy production constraints (Navier--Stokes)}
\label{sec:EPCNS}

In the TH model, the NS entropy production \EQn{eq:divSNS2} equals
\begin{equation}
\p_\mu S^\mu_{\rm NS} = - 2 \, \dvv \cdot \Cv_k
+ 2  \evv \cdot \Cv_\omega 
\end{equation}
and is positive if
\begin{equation}
\dvv = - 2 \lambda T \, \Cv_k, \quad \evv = \gamma  T \, \Cv_\omega,
\label{eq:dbiutNS1}
\end{equation}
which is consistent with \EQSTWO{eq:daTH}{eq:eaTH}. It is readily seen that~\EQS{eq:dbiutNS1} represent additional constraints on the system of Eqs.~\EQn{eq:BNcon}, \EQSMn{eq:E0TH}{eq:E3TH}, and~\EQn{eq:cs}. With \EQS{eq:dbiutNS1} included, the system of Eqs.~\EQn{eq:BNcon}, \EQSMn{eq:E0TH}{eq:E3TH}, and~\EQn{eq:cs} becomes overdetermined -- we obtain seventeen equations for fourteen unknowns.

\begin{figure}[t]
    \centering
\includegraphics[width=0.5\columnwidth]{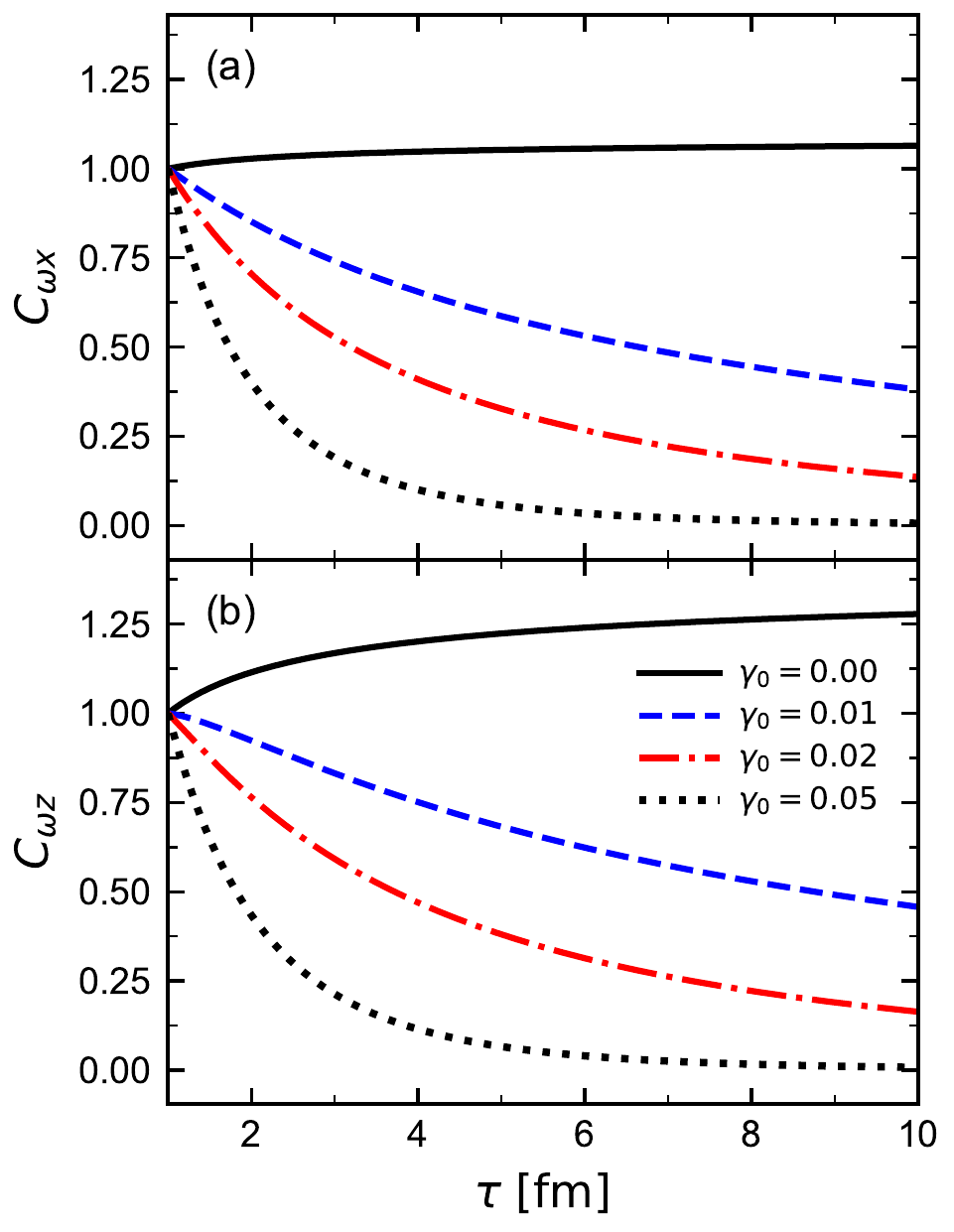}
\caption{Proper-time evolution of (a) transverse spin polarization $C_{\omega x}(\tau)$ and (b) longitudinal spin polarization $C_{\omega z}(\tau)$. The curves show the relaxation dynamics for different values of the dimensionless kinetic coefficients $\gamma_0 \in \{0.00,\, 0.01,\, 0.02,\, 0.05\}$ with initial conditions $C_{\omega x}(\tau_0) = C_{\omega z}(\tau_0) = 1$.}
    \label{fig:th_ns}
\end{figure}

The origin of this difficulty is the appearance of the vector $\dvv$ in the equations of motion~\EQSMn{eq:E0TH}{eq:E3TH}. If the components of $\dvv$ were neglected in~\EQSMn{eq:E0TH}{eq:E3TH}, we would have equal numbers of equations and unknown functions. Neglecting $\dvv$ in~\EQSMn{eq:E0TH}{eq:E3TH} is, however, inconsistent with keeping it on the right-hand side of \EQ{eq:cs}. These arguments motivate us to conclude that the functions $\Cv_k$ and, consequently, $\dvv$ should be set equal to zero. This assumption is consistent with the fact that thermal vorticity is zero and the inclusion of dissipation should result in $\omega$ approaching $\varpi =0$.

It remains for us to explain the behavior in the magnetic sector described by the vectors $\evv$ and $\Cv_\omega$. This is done numerically. The coupled system of ordinary differential equations is solved by employing a Runge--Kutta integration scheme. The initial proper time for the hydrodynamic evolution is set to $\tau_0 = 1.0$~fm. The hydrodynamic background is initialized with the temperature $T_0 = 0.155$~GeV and the baryon chemical potential $\mu_0 =0.8$~GeV. The particle mass is set to $m = 1.116$~GeV, which corresponds to the $\Lambda$ hyperon. The initial conditions for the spin polarization components are fixed uniformly at $C_{\omega x}(\tau_0) = C_{\omega y}(\tau_0) = C_{\omega z}(\tau_0) = 1$. The numerical integration is performed up to $\tau_f = 10.0$~fm. The resulting evolution is presented in \FIG{fig:th_ns}. We explicitly assume that the spin kinetic coefficient $\gamma$ is proportional to the entropy density, using the form $\gamma = \gamma_0 \sigma$. The curves demonstrate the relaxation dynamics governed by different values of the dimensionless parameter $\gamma_0 \in \{0.00, 0.01, 0.02, 0.05\}$. 

In the nondissipative case ($\gamma_0 = 0.00$), the spin polarization continuously increases. This behavior has been already found in the studies of perfect fluid hydrodynamics~\CITn{Florkowski:2019qdp}. The inclusion of the entropy production constraints introduces damping effect on the spin polarization. This damping becomes stronger for larger values of $\gamma_0$.

\subsection{Entropy production constraints (Israel--Stewart)}
\label{sec:EPCIS}

In the IS case and TH geometry, \EQSTWO{eq:IS_d_gen}{eq:IS_e_gen} become two differential vector equations:
\begin{equation}
\tau_d \, {\dot \dvv} + \dvv = - \lambda T \, \LB 2 \Cv_k + {\dot a_4} \, \dvv + \f{a_4}{\tau} \, \dvv \RB
\label{eq:TH_IS_dv}
\end{equation}
and
\begin{equation}
\tau_e \, {\dot \evv} + \evv = 
\gamma T \, \LB \Cv_\omega + {\dot a_5} \, \evv + \f{a_5}{\tau} \, \evv \RB.
\label{eq:TH_IS_de}
\end{equation}
Similarly to the NS case, we observe that the system is overdetermined, unless the dissipation vector $\dvv$ and the related vector $\Cv_k$ vanish. Consequently, we set $\dvv = \Cv_k = 0$ and consider only the evolution of the vector $\evv$ governed by~\EQ{eq:TH_IS_de}.

\begin{figure}[t]
\centering \includegraphics[width=0.85\textwidth]{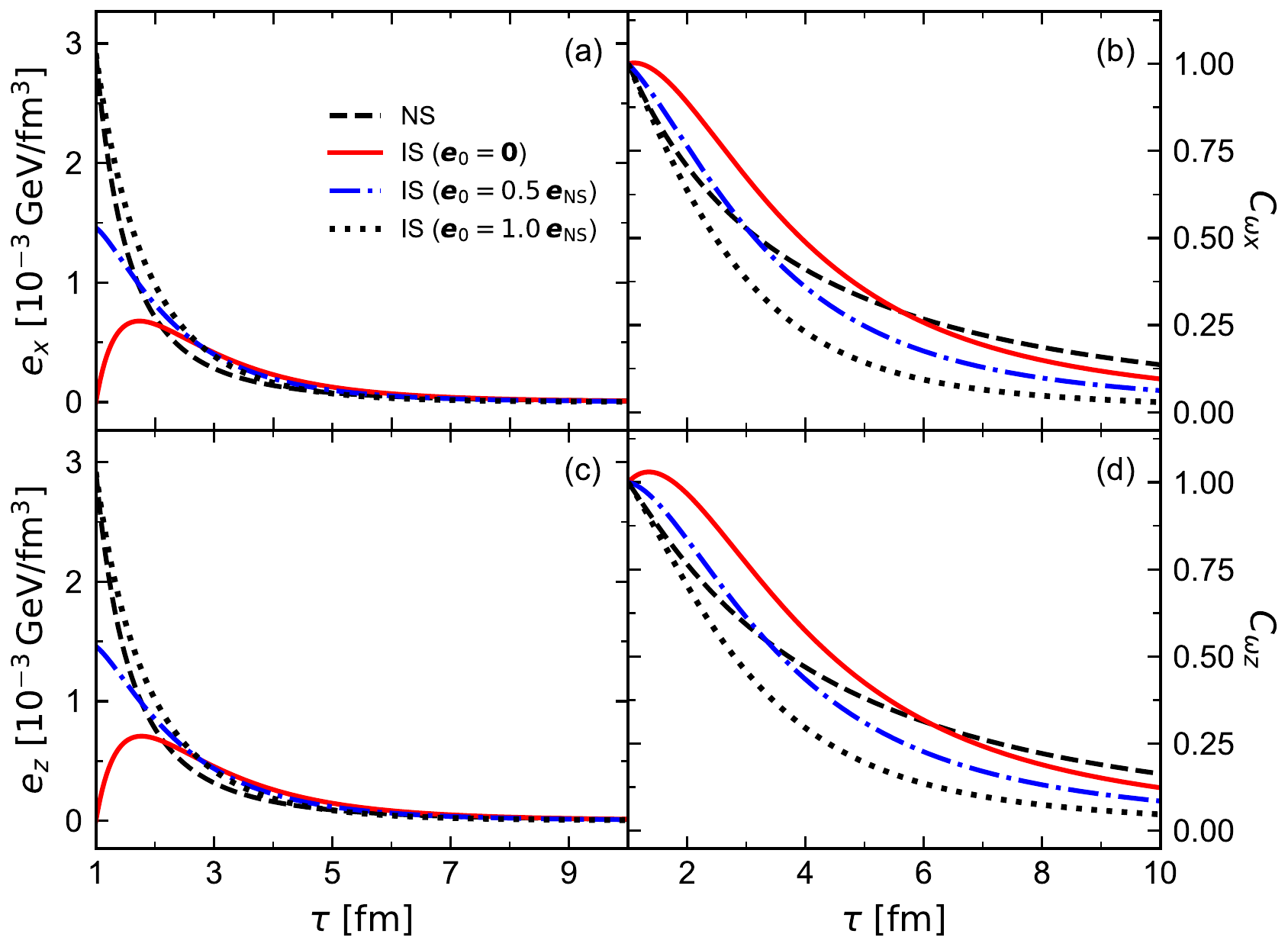}
\caption{Proper-time evolution of the components of the magnetic dissipative vector $\evv$ and the components of the spin polarization vector $\Cv_\omega$ in the Israel--Stewart formulation with $\tau_e = 1.0$~fm, compared with the Navier--Stokes asymptotic limit ($\bm{e} = \gamma T \bm{C}_\omega$). The panels show (a) the transverse component $e_x(\tau)$, (b) the transverse spin polarization component $C_{\omega x}(\tau)$, (c) the longitudinal component $e_z(\tau)$, and (d) the longitudinal spin polarization component $C_{\omega z}(\tau)$. Different line styles correspond to different initial dissipation magnitudes of $\evv_0$, as described in the legend.}
    \label{fig:th_is}
\end{figure}


The numerical integration scheme and the thermodynamic background parameters remain identical to those used in \SEC{sec:EPCNS}. To examine the dynamics of the dissipation vector $\evv$,  we first set the relaxation time to $\tau_e = 1.0$~fm and vary the initial condition $\evv_0 = \bm{e}(\tau_0)$. We explore three distinct scenarios: $\evv_0 = \bm{0}$, $\evv_0 = 0.5~\bm{e}_{\mathrm{NS}}$, and $\evv_0  = 1.0~\bm{e}_{\mathrm{NS}}$. Here, $\bm{e}_{\mathrm{NS}} = \gamma(\tau_0) T(\tau_0) \bm{C}_\omega(\tau_0)$ denotes the initial dissipation vector predicted by the Navier--Stokes limit. The inclusion of an independent dynamic equation for $\evv(\tau)$ leads to the behavior depicted in \FIG{fig:th_is}, which shows the proper time evolution of the components $e_x(\tau)$, $e_z(\tau)$, $C_{\omega \,x}(\tau)$, and $C_{\omega \,z}(\tau)$. Due to the transverse homogeneity, the evolutions of the $x$ and $y$ components are the same.

\begin{figure}[t]
    \centering
\includegraphics[width=0.85\textwidth]{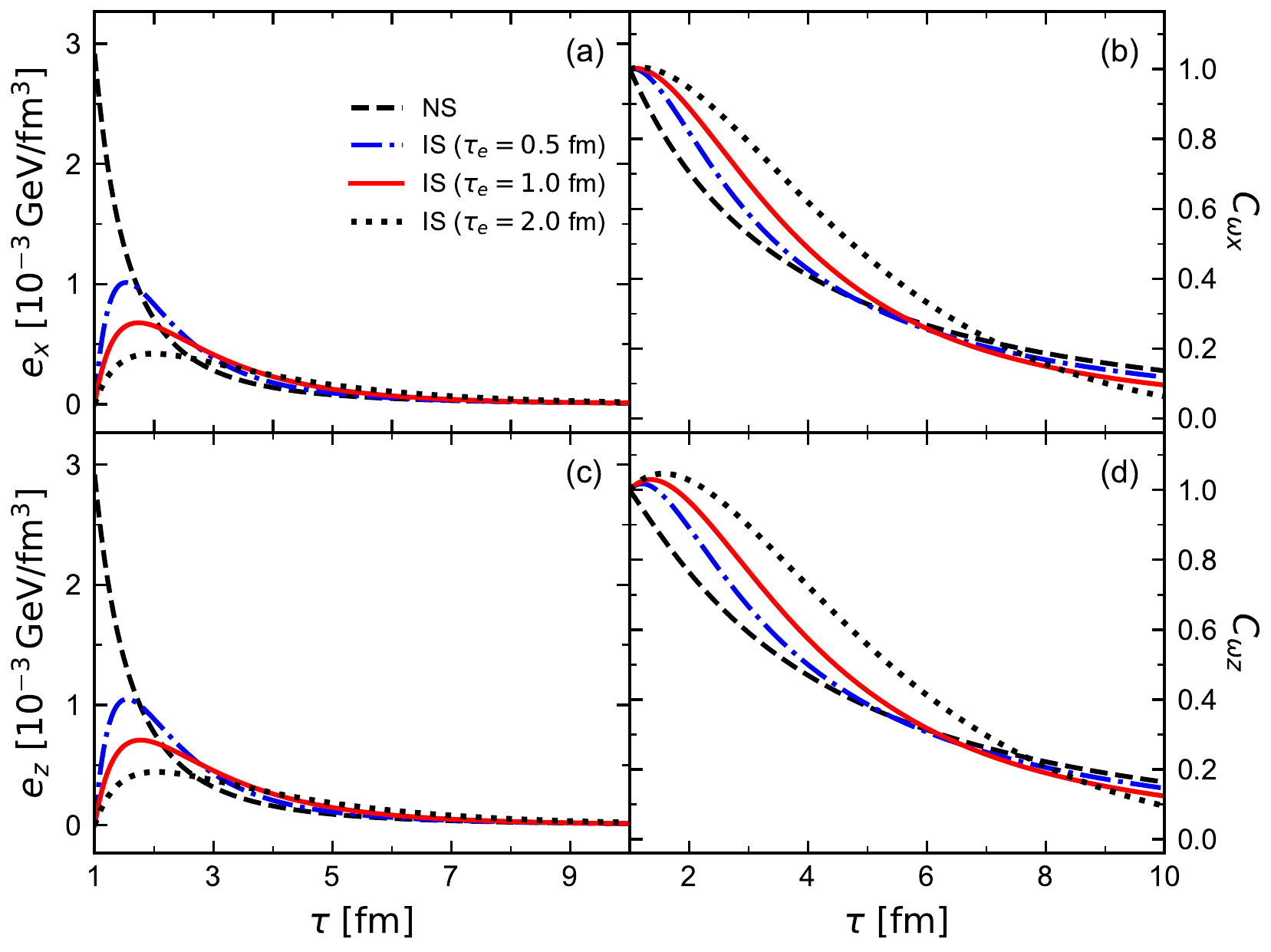}
    \caption{Proper-time evolution of the dissipation and spin polarization vectors in the Israel--Stewart formulation for different relaxation times, $\tau_e \in \{0.5,\, 1.0,\, 2.0\}$~fm. The system is initialized at $\bm{e}_0 = \bm{0}$ with $\bm{C}_{\omega}(\tau_0) = 1$. The panels display (a) transverse dissipation $e_x$, (b) transverse spin polarization $C_{\omega x}$ (with $C_{\omega y} \equiv C_{\omega x}$), (c) longitudinal dissipation $e_z$, and (d) longitudinal spin polarization $C_{\omega z}$. All results obtained in Israel--Stewart approach are compared to the Navier--Stokes limit (black dashed lines).}
    \label{fig:is_tau_e0}
\end{figure}

If the system is initialized with zero anisotropy ($\evv_0 = \bm{0}$), the dissipative vector  requires certain time to develop. This delay implies that the initial decay of the spin polarization is noticeably slower than in the Navier--Stokes limit  -- as long as the right-hand side of \EQn{eq:cs} is negligible, the spin evolution is governed by the perfect-fluid equations. In contrast, if the system is initialized with the fully developed dissipation vector ($\bm{e}_0 = 1.0~\bm{e}_{\mathrm{NS}}$), the spin polarization is subject to strong damping. Regardless of the chosen initial state, all the Israel--Stewart cases converge towards the Navier--Stokes result. This behavior is similar to that found in standard relativistic hydrodynamics, where the IS dynamics reaches local equilibrium by first approaching the NS regime. 

\begin{figure}[t]
\centering    \includegraphics[width=0.85\textwidth]{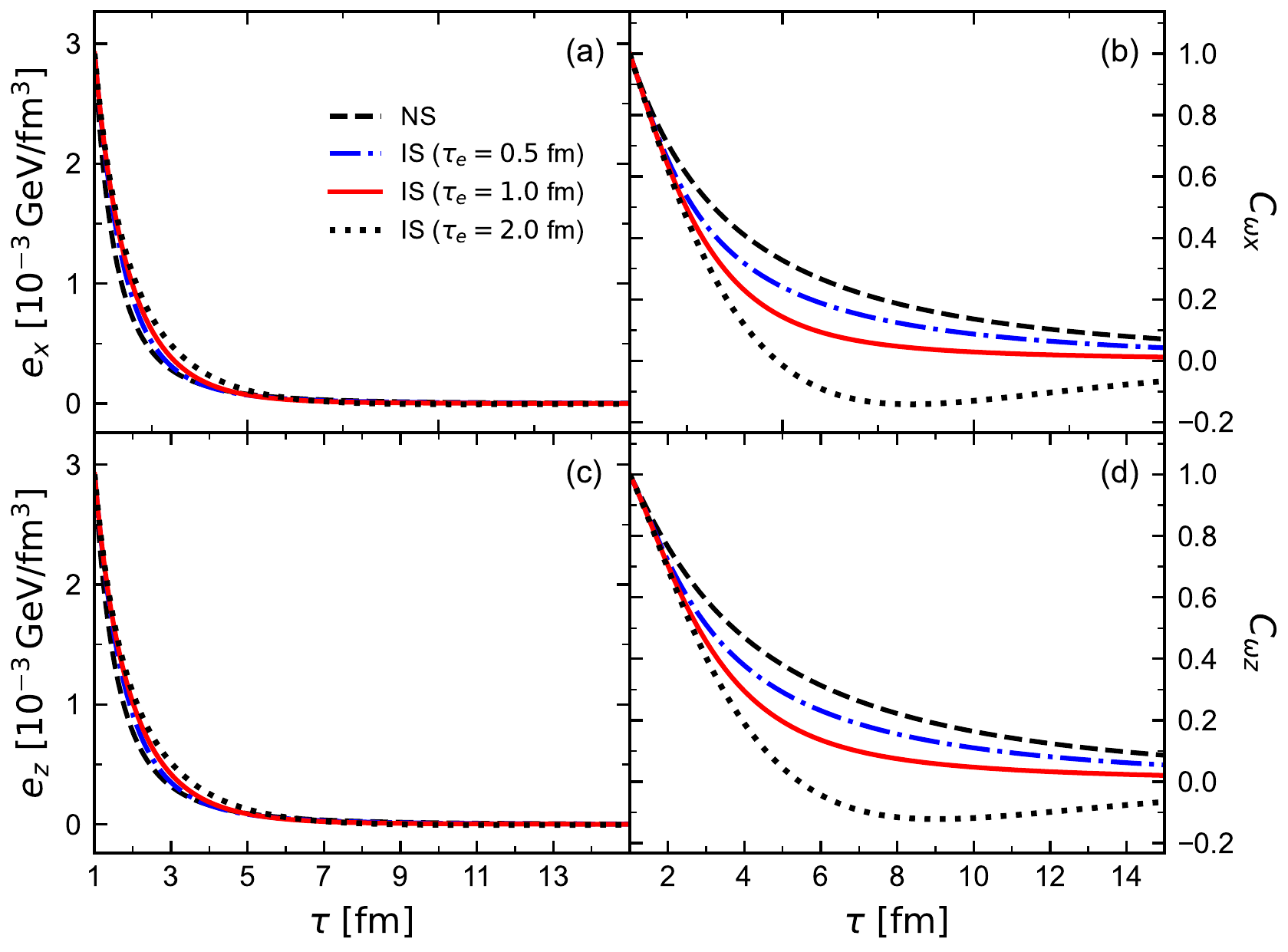}
\caption{Proper-time evolution of the dissipation and spin polarization vectors for the case $\bm{e}(\tau_0) = \bm{e}_{\mathrm{NS}}(\tau_0)$ and different relaxation times, $\tau_e \in \{0.5, 1.0, 2.0\}$~fm. The final evolution time is $\tau_f = 15.0$~fm. The panels display (a) transverse dissipation $e_x(\tau)$, (b) transverse spin polarization $C_{\omega x}(\tau)$, (c) longitudinal dissipation $e_z(\tau)$, and (d) longitudinal spin polarization $C_{\omega z}(\tau)$. }
    \label{fig:is_tau_test_ns_15}
\end{figure}

To investigate the role of the relaxation time in the Israel--Stewart formulation, we perform computations for $\tau_e = 0.5$~fm, $\tau_e = 1.0$~fm, and $\tau_e = 2.0$~fm, assuming  $\evv_0 = 0$. The corresponding evolution of the dissipation and spin polarization vector components is presented in \FIG{fig:is_tau_e0}. The relaxation time determines how quickly the dissipation vector components develop and approach the NS limit. For larger values of $\tau_e$, the dissipation vector requires more time to reach this limit.

Our results for a scenario where the system is initialized in the NS regime, $\bm{e}(\tau_0) = \bm{e}_{\mathrm{NS}}(\tau_0)$, are shown in~\FIG{fig:is_tau_test_ns_15}. To observe the full relaxation process, numerical integration has been extended to the final time $\tau_f = 15.0$~fm. We observe in this case that the IS evolution of $\evv(\tau)$ follows closely the NS dynamics, although small differences can be seen. Larger differences between the IS and NS results are seen for the spin polarization. As expected, for larger relaxation times the IS dynamics approaches the NS limit later. Interestingly, for $\tau_e = 2.0$~fm the components of the spin polarization vector (being positive initially) change sign during the time evolution and approach zero from below -- the corresponding NS values approach zero from above.

\section{Boost-invariant and cylindrically symmetric systems}
\label{sec:CS}

We turn now to a discussion of the system that is boost-invariant and cylindrically symmetric in the transverse plane. A novel feature, compared to the TH model, is the inclusion of the radial flow. We call this type of expansion shortly cylindrically symmetric (CS). We follow essentially the same steps as in the discussion of the TH model. The basic difference is in the more complicated tensor structures present in the CS formulation. A version of the CS model featuring a perfect spin fluid was recently investigated in \CITn{Drogosz:2026egr}; the present work can therefore be viewed as a direct extension of \CITn{Drogosz:2026egr} that incorporates dissipative phenomena.
\subsection{Basis vectors and basic symmetries}
\label{sec:bicsbasis}

\begin{figure}[t]
\centering
\subfloat[]{\includegraphics[width=0.5\textwidth]{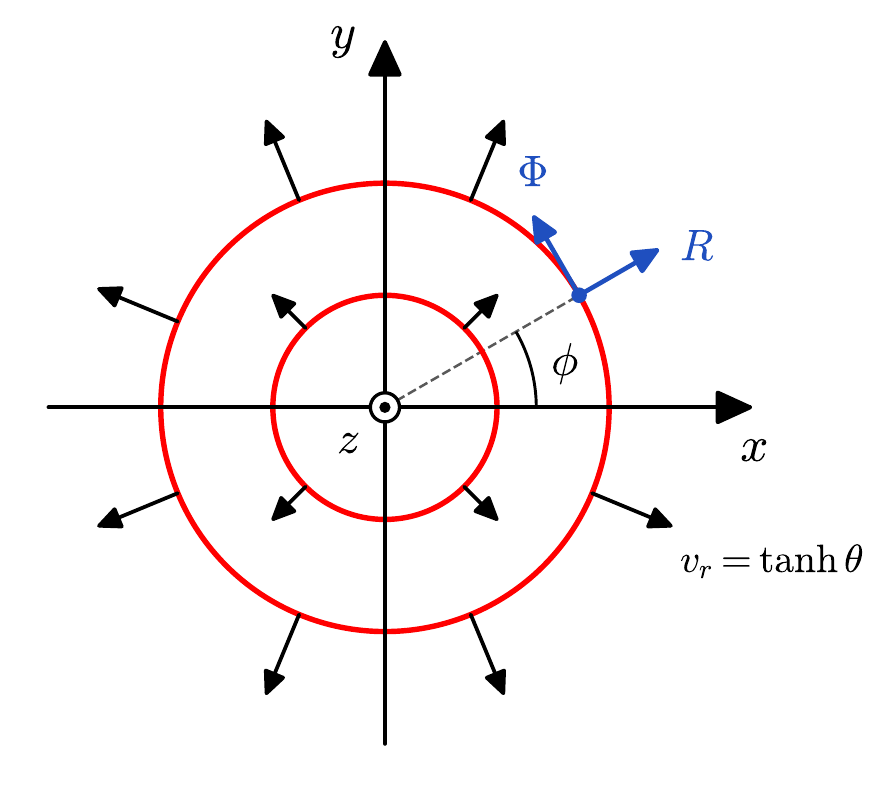}\label{fig:cs_geometry_a}}
\hfill
\subfloat[]{\includegraphics[width=0.45\textwidth]{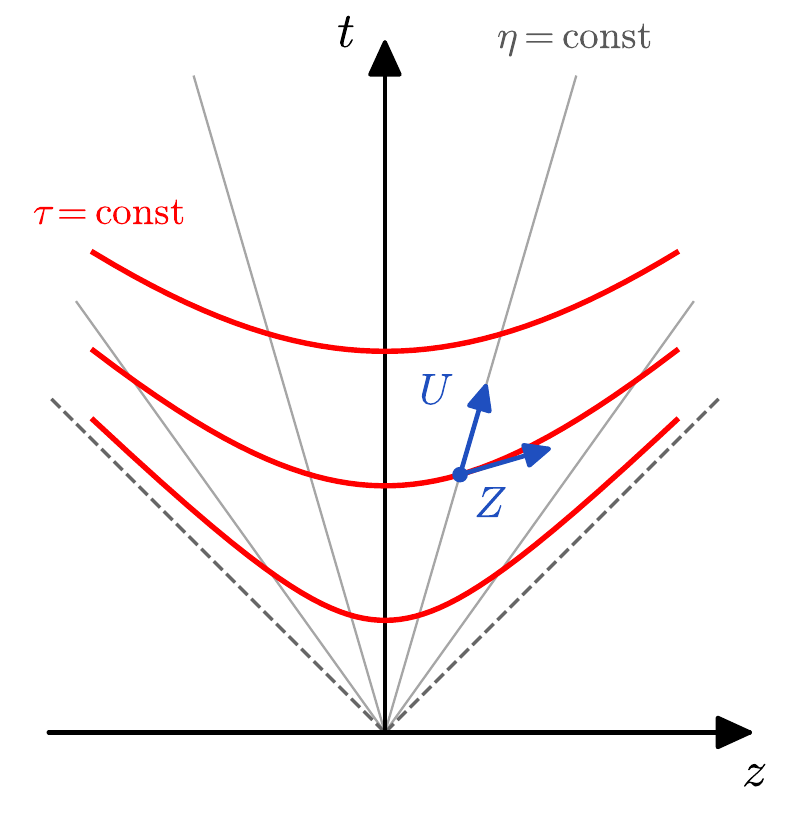}\label{fig:cs_geometry_b}}
\caption{Schematic view of the CS geometry. (a) Transverse plane $z = 0$: the red circles denote surfaces of constant $r$, the black arrows indicate the radial flow with the transverse velocity $v_r = \tanh\theta$, and the blue arrows show the basis vectors $R$ and $\Phi$, \EQSTWO{eq:CSRmu}{eq:CSPhimu}, at a point with the azimuthal angle $\phi$; the $z$ axis points out of the plane. (b) Longitudinal $(z, t)$ plane at a fixed transverse position: hyperbolas of constant proper time $\tau$ (red), lines of constant spacetime rapidity $\eta$ (gray), and the light cone (dashed). The basis vector $U$ is tangent to the line $\eta = {\rm const}$, while $Z$, \EQ{eq:CSZmu}, is tangent to the hyperbola $\tau = {\rm const}$. }
    \label{fig:cs_geometry}
\end{figure}

The boost-invariant and cylindrically symmetric basis ${\cal X}^\mu_{(i)} \, (i=0,1,2,3)$ consists now of the vectors $U, R, \Phi$, and $Z$ defined by the following expressions~\CITn{Florkowski:2011jg}:
\begin{eqnarray}
{\cal X}^\mu_{(0)} = U^\mu &=& \LB \cosh \theta \cosh \eta,  \sinh \theta \cos  \phi,            \sinh \theta \sin  \phi, 
\cosh \theta \sinh \eta \RB,
\label{eq:CSUmu}
\end{eqnarray}
\begin{eqnarray}
{\cal X}^\mu_{(1)} = R^\mu &=& \LB
\sinh \theta \cosh \eta, 
\cosh \theta \cos  \phi,          
\cosh \theta \sin  \phi,           \sinh \theta \sinh \eta \RB,
\label{eq:CSRmu}
\end{eqnarray}
\begin{eqnarray}
{\cal X}^\mu_{(2)} = \Phi^\mu &=& \LB
0,
-\sin  \phi, 
\cos  \phi, 0 \RB,
\label{eq:CSPhimu}
\end{eqnarray}
and
\begin{eqnarray}
{\cal X}^\mu_{(3)} = Z^\mu &=& \LB
\sinh \eta,
0,
0,
\cosh \eta \RB.
\label{eq:CSZmu}
\end{eqnarray}
See \FIG{fig:cs_geometry} for a schematic view. Here $r$ and $\phi$ are polar coordinates in the transverse plane (radius and azimuthal angle), while $\theta = \theta(\tau,r)$ is the transverse rapidity. At $z=0\, (\eta=0)$ the transverse velocity $v_r(\tau,r) = \tanh\theta(\tau,r)$. Similarly to~\EQn{eq:normTH}, we have
\begin{eqnarray}
U^2 &=& 1, \quad Z^2 = R^2 = \Phi^2 = -1, \nonumber \\
U \cdot Z &=& 0, \quad U \cdot R = 0, \quad U \cdot \Phi = 0, \nonumber \\
Z \cdot R &=& 0, \quad R \cdot \Phi = 0, \quad \Phi \cdot Z = 0.
\label{eq:normCS}
\end{eqnarray}
The analog of the identities~\EQn{eq:basisLC} is the following set of equations
\begin{align}\begin{split}
\epsilon_{\alpha\beta\gamma\delta} U^\beta R^\gamma \Phi^\delta = - Z_\alpha, \\
\epsilon_{\alpha\beta\gamma\delta} U^\beta \Phi^\gamma Z^\delta = - R_\alpha, \\
\epsilon_{\alpha\beta\gamma\delta} U^\beta Z^\gamma R^\delta = - \Phi_\alpha. \\
\label{eq:basisLCCS}
\end{split}\end{align}

In the TH model we deal with scalar functions of the proper time. In the CS model, scalar functions depend on the proper time and the distance $r$ from the symmetry axis. For such functions, the total time derivative equals 
\begin{eqnarray}
D f(\tau,r) \equiv U^\mu \p_\mu f(\tau,r) &=& 
\LB\cosh\theta \,\p_\tau + \sinh\theta\, \p_r \RB f\LB\tau, r\RB \nn \\
&=& 
\LB\cosh\theta \,{\dot f\LB\tau, r\RB} 
+ \sinh\theta\, f^\prime\LB\tau, r\RB \RB .
\label{eq:DfCS}
\end{eqnarray}
Here we have introduced a dot (a prime) as the symbol of the partial derivative with respect to proper time (radius coordinate). Similarly, we define the transverse radial derivative
\begin{eqnarray}
\nabla_R f\LB\tau, r\RB \equiv R^\mu \p_\mu f\LB\tau, r\RB = \LB \sinh \theta \,{\dot f\LB\tau, r\RB}  
+ \cosh\theta\, f^\prime\LB\tau, r\RB \RB .
\label{eq:nRfCS}
\end{eqnarray}
One can notice that the operators $D$ and $\nabla_R$ differ by exchange of the hyperbolic functions. The transverse gradient of a scalar function can be written as
\begin{equation}
\nabla^\mu f(\tau,r) = \Delta^{\mu \nu} \p_\nu f(\tau,r) = - \LB \sinh \theta \,{\dot f\LB\tau, r\RB}  
+ \cosh\theta\, f^\prime\LB\tau, r\RB \RB R^\mu =
- \nabla_R f\LB\tau, r\RB  R^\mu.
\label{eq:nablafCS}
\end{equation}
Furthermore, for the CS model one finds:
\begin{eqnarray}
a^\mu = \LB \cosh\theta\, \dot \theta +\sinh\theta\, \theta' \RB R^\mu &=& D\theta(\tau,r) R^\mu \equiv a_r(\tau,r) R^\mu,  \\
\nabla^{[ \mu} U^{\nu ]} &=& 0.
\label{eq:CScond}
\end{eqnarray}
Using \EQSTWO{eq:nablafCS}{eq:CScond}, one finds that the thermal vorticity $\varpi_{\mu\nu}$ has the structure of \EQ{eq:o1} with the vanishing magneticlike
part, namely
\bea
\varpi_{\mu \nu}=
k^{\rm eq}_{\mu} \, U_{\nu} -k^{\rm eq}_{\nu} \,U_{\mu} =  C^{\rm eq}_{kr} \LB R_\mu U_\nu - R_\nu U_\mu \RB
\label{eq:to1}
\eea
where
\begin{equation}
C^{\rm eq}_{kr} = \f{\beta}{2} \LB  D\theta - \beta \nabla_R T \RB = \f{\beta}{2} \LB a_r - \beta \nabla_R T \RB .
\label{eq:varpiCS}
\end{equation}
Following~\CITn{Florkowski:2019qdp}, we denote the coefficients of thermal vorticity by the superscript ``eq'', since they define the values of the coefficients of the spin polarization tensor at which the entropy production vanishes.

For the spin polarization tensor $\omega$, we use now a parametrization whose form is the same as \EQ{eq:o1}, however, the vectors $k$ and $\omega$ are expanded in the basis \EQn{eq:CSRmu}--\EQn{eq:CSZmu}, 
\begin{align}
k^{\mu} = C_{kr}R^{\mu} + C_{k\phi}\Phi^{\mu} + C_{kz}Z^{\mu}, \quad
\omega^{\mu} = C_{\omega r}R^{\mu} + C_{\omega \phi}\Phi^{\mu} + C_{\omega z}Z^{\mu} .
\label{eq:CkCoCS}
\end{align}
Similarly, the dissipative vector $d^\mu$ and the antisymmetric tensor $e_a^{\mu\nu}$ can be decomposed as
\begin{eqnarray}
d^\mu &=& d_r R^\mu + d_\phi \Phi^\mu + d_z Z^\mu, \label{eq:bics_dmu} \\
e_a^{\mu\nu} &=& -e_r (\Phi^\mu Z^\nu - \Phi^\nu Z^\mu) -e_\phi (Z^\mu R^\nu - Z^\nu R^\mu) - e_z (R^\mu \Phi^\nu - R^\nu \Phi^\mu). \label{eq:bics_emu}
\end{eqnarray}
Below, following~\EQn{eq:3vecnot}, we use the three-vector notation
\begin{eqnarray}
\Cv_k = \LB C_{kr}, C_{k\phi},  C_{kz} \RB,&& \quad
\Cv_\omega = \LB C_{\omega r}, C_{\omega \phi}, C_{\omega z} \RB,
\label{eq:3vecnotCS1}
\end{eqnarray}
\begin{eqnarray}
\dvv = \LB d_r, d_\phi,  d_z \RB,&& \quad
\evv = \LB e_r, e_\phi,  e_z \RB.
\label{eq:3vecnotCS2}
\end{eqnarray}
We note that all the scalar coefficients depend exclusively on proper time $\tau$ and radial coordinate~$r$. Similarly as in the TH case, we call the vectors $\dvv$ and $\evv$ the electric and magnetic dissipation vectors.

\subsection{Conservation laws}
\label{sec:bicscon}

In this section we present the explicit forms of all the conservation equations used in the CS~model.

\subsubsection{Baryon number conservation}
\label{sec:bicsNcon}

Implementation of boost invariance and cylindrical symmetry in the baryon conservation equation $\p_\mu (n U^\mu) = 0$ gives
\begin{eqnarray}
\left( \frac{\partial}{\partial \tau} 
+ \tanh \theta \frac{\partial}{\partial r} \right) n + n \left[  \left( \frac{1}{\tau} 
+ \frac{\partial \theta}{\partial r} \right) + \tanh \theta \left( 
\frac{1}{r} + \frac{\partial \theta}{\partial \tau} \right) \right] =0.  \label{eq:ncon}
\end{eqnarray}
In the limit $\theta \to 0$, \EQn{eq:ncon} reduces to \EQn{eq:BNcon}.

\subsubsection{Energy--momentum conservation}
\label{sec:bicsEMcon}

Projecting \EQ{eq:enmomcon} onto the flow vector $U_\nu$, we obtain the energy evolution equation
\begin{align}
D\epsilon + \Theta (\epsilon + P) + \nabla_R d_r + d_r \left(\frac{\sinh \theta}{\tau} + \frac{\cosh \theta}{r} \right) = 0.
\label{eq:E0CS}
\end{align}
Similarly, contraction of~\EQ{eq:enmomcon} with the radial vector $R_\nu$ yields the momentum conservation equation in the transverse plane
\begin{align}
\nabla_R P + (\epsilon + P) D\theta -D d_r - d_r \left( \frac{\cosh \theta}{\tau} + \frac{\sinh \theta}{r} \right) = 0.
\label{eq:E1CS}
\end{align}
The projections onto the azimuthal vector $\Phi_\nu$ and the longitudinal vector $Z_\nu$ receive no contributions from the symmetric equilibrium tensor $T^{\mu\nu}_{\rm eq}$, isolating the dynamics of the transverse dissipative degrees of freedom. For the azimuthal projection, we find
\begin{align}
Dd_\phi + d_\phi \left(\nabla_R \theta + \frac{\cosh \theta}{\tau} \right) + \nabla_R e_z + e_z \left(D\theta + \frac{\sinh \theta}{\tau} \right) = 0,
\label{eq:E2CS}
\end{align}
whereas for the longitudinal projection we obtain
\begin{align}
Dd_z + d_z \left(\nabla_R \theta + \frac{\sinh \theta}{r} \right) -\nabla_R e_\phi - e_\phi \left(D\theta + \frac{\cosh \theta}{r} \right) = 0.
\label{eq:E3CS}
\end{align}

It is interesting to compare \EQSM{eq:E0CS}{eq:E3CS} with \EQSM{eq:E0TH}{eq:E3TH}. Three out of four equations in \EQSMn{eq:E0TH}{eq:E3TH} can be interpreted as constraints on spin dynamics. In the CS model, only two out of four equations in \EQSMn{eq:E0CS}{eq:E3CS} play such a role. We thus see that a more developed flow pattern restricts the spin dynamics less. 

In the CS model, we also observe that the dissipation terms appear in \EQSTWO{eq:E0CS}{eq:E1CS}. This means that we cannot separate a spinless hydrodynamic background from spin evolution, unless the terms involving $d_r$ are numerically negligible in \EQSTWO{eq:E0CS}{eq:E1CS}.

\subsubsection{Angular momentum}

\noindent
{\bf (a)} The radial electriclike sector obtained from the contraction of Eq.~(\ref{eq:conNSspin}) with $U_\mu R_\nu$,
\begin{equation}
D(A C_{kr}) + A C_{kr} \left( \Theta + \frac{\cosh\theta}{2\tau} + \frac{\sinh\theta}{2r} \right) = -2d_r.
\label{eq:SaCS}
\end{equation}

\noindent 
{\bf (b)} The radial magneticlike sector obtained from the contraction of Eq.~(\ref{eq:conNSspin}) with $\Phi_\mu Z_\nu$,
\begin{equation}
D(A_1 C_{\omega r}) + C_{\omega r} \left[ A_1 \Theta - \frac{A}{2} \left( \frac{\cosh\theta}{\tau} + \frac{\sinh\theta}{r} \right) \right] = - 2e_r.
\end{equation}

\noindent
{\bf (c)} The longitudinal magneticlike sector obtained from the contractions of Eq.~(\ref{eq:conNSspin}) with $U_\mu \Phi_\nu$ and $R_\mu \Phi_\nu$,
\begin{align}
&D(A C_{k\phi})\!-\!\frac{1}{2} \nabla_R (A C_{\omega z})\!+\!A \left[ \frac{3}{2} C_{k\phi} \left( \Theta\!-\!\frac{\sinh\theta}{3r} \right) - \frac{C_{\omega z}}{2} \left( D\theta\!+\! \frac{\sinh\theta}{\tau} \right) \right]\!+\!A_1 C_{\omega z} D\theta\!+\!2d_\phi = 0 , \label{eq:CSc1} \\
&D(A_1 C_{\omega z})\!+\!\frac{1}{2} \nabla_R (A C_{k\phi})\!+\!A \left[ \frac{3}{2} C_{k\phi} \left( D\theta\!+\!\frac{\cosh\theta}{3r} \right) - \frac{C_{\omega z}}{2} \left( \Theta\!-\! \frac{\cosh\theta}{\tau} \right) \right]\!+\!A_1 C_{\omega z} \Theta + 2 e_z = 0 .
\label{eq:CSc2}
\end{align}

\noindent
{\bf (d)} The longitudinal electriclike sector obtained from the contractions of Eq.~(\ref{eq:conNSspin}) with $U_\mu Z_\nu$ and $Z_\mu R_\nu$,

\begin{align}
&D(A C_{kz})\!+\!\frac{1}{2} \nabla_R (A C_{\omega\phi})\!+\!A \left[ \frac{3}{2} C_{kz} \left( \Theta\!-\!\frac{\cosh\theta}{3\tau} \right) + \frac{C_{\omega\phi}}{2} \left( D\theta\!+\! \frac{\cosh\theta}{r} \right) \right]\!-\!A_1 C_{\omega\phi} D\theta\!+\!2d_z = 0 \label{eq:CSd1} , \\
&D(A_1 C_{\omega\phi})\!-\!\frac{1}{2} \nabla_R (A C_{kz})\!-\!A \left[ \frac{3}{2} C_{kz} \left( D\theta\!+\!\frac{\sinh\theta}{3\tau} \right)\!+\! \frac{C_{\omega\phi}}{2} \left( \Theta\!-\! \frac{\sinh\theta}{r} \right) \right]\!+\!A_1 C_{\omega\phi} \Theta\!+\!2e_\phi = 0 . \label{eq:CSd2} 
\end{align}

\subsection{Entropy production constraints (Navier--Stokes)}

Using \EQSTWO{eq:nablafCS}{eq:CScond} and the decompositions \EQn{eq:3vecnotCS1}, \EQn{eq:bics_dmu}, and \EQn{eq:bics_emu}, we find that in the CS model the NS entropy production \EQn{eq:divSNS2} equals
\begin{equation}
\p_\mu S^\mu_{\rm NS} = d_r \LB \beta D\theta - \beta^2 \nabla_R T \RB - 2 \, \dvv \cdot \Cv_k + 2 \evv \cdot \Cv_\omega
\label{eq:CS_divS1}
\end{equation}
and is positive if
\begin{gather}
d_r = \lambda T \LB \beta D\theta - \beta^2 \nabla_R T - 2 C_{kr} \RB, \quad 
d_\phi = - 2 \lambda T \, C_{k\phi}, \quad
d_z = - 2 \lambda T \, C_{kz},
\label{eq:CS_NS_dr} \\
\evv = \gamma T \, \Cv_\omega,
\label{eq:CS_NS_e}
\end{gather}
which is consistent with \EQSTWO{eq:da}{eq:ea}. Equations \EQn{eq:CS_NS_dr} and \EQn{eq:CS_NS_e} are the CS analogs of \EQS{eq:dbiutNS1}. The only difference is the appearance of the terms $\beta D\theta$ and $-\beta^2 \nabla_R T$ in the expression for $d_r$. They originate from thermal vorticity \EQn{eq:varpi}, which, in contrast to the TH model, does not vanish in the CS geometry. Thus the first equation in \EQn{eq:CS_NS_dr} can be written as $d_r = 2 \lambda T (C^{\rm eq}_{kr} - C_{kr})$, and the entropy production \EQn{eq:CS_divS1} becomes
\begin{equation}
\p_\mu S^\mu_{\rm NS} = 4 \lambda T \LSB \LB C^{\rm eq}_{kr} - C_{kr} \RB^2 + C_{k\phi}^2 + C_{kz}^2 \RSB
+ 2 \gamma T \, \Cv_\omega \cdot \Cv_\omega \geq 0 .
\label{eq:CS_divS}
\end{equation}
Consequently, the entropy production vanishes if $C_{kr} = C^{\rm eq}_{kr}$, while all other coefficients of the spin polarization tensor vanish. Hence, in the CS model we expect that dissipation drives the radial electriclike coefficient $C_{kr}$ towards $C^{\rm eq}_{kr}$, while the other components approach zero or should be equal to zero.

In fact, if the functions $T$, $\mu$, and $\theta$ are known, \EQSTWO{eq:CSc1}{eq:CSc2} together with \EQ{eq:E2CS} become three equations for four unknowns: $C_{k\phi}$, $C_{\omega z}$, $d_\phi$, and $e_z$. Including the NS conditions connecting $d_\phi$ with $C_{k\phi}$ and $e_z$ with $C_{\omega z}$, we obtain five equations for four unknowns, i.e., an overdetermined system. A natural solution in this case is to set all these functions equal to zero. Similar analysis of \EQSTWO{eq:CSd1}{eq:CSd2} together with \EQ{eq:E3CS} leads us to the conclusion that $C_{kz}$, $C_{\omega \phi}$, $d_z$, and $e_\phi$ are zero. Thus, only radial components $d_r$ and $e_r$ (or $C_{kr}$ and $C_{\omega r}$) exhibit a nontrivial dynamics. 

One more important remark should be made about the presence of the time derivative of $d_r$ in \EQSTWO{eq:E0CS}{eq:E1CS}. In the NS approach, this derivative should be obtained by differentiation of \EQ{eq:CS_NS_dr}, which introduces second order proper-time derivatives of $\theta$ and $T$. Such a procedure spoils the structure of hydrodynamic equations, indicating strong limitations of the NS approach in spin physics. Due to such issues, we do not perform a numerical analysis of the NS case. We show below that a natural solution to this problem is to use the IS approach.

\subsection{Entropy production constraints (Israel--Stewart)}

In the CS system, the expansion scalar is given by $\Theta \equiv \partial_\mu U^\mu = \cosh\theta/\tau + \sinh\theta/r + \nabla_R\theta$, with $\nabla_R\theta = \sinh\theta\,\dot\theta + \cosh\theta\,\theta'$. By projecting \EQSTWO{eq:IS_d_gen}{eq:IS_e_gen} onto the basis vectors $R^\mu$, $\Phi^\mu$, and $Z^\mu$, and using the relations $D R^\mu = D\theta\, U^\mu$ and $D \Phi^\mu = D Z^\mu = 0$ (see App.~\ref{sec:BICSuseful}), we obtain two differential equations for the vectors $\dvv$ and $\evv$ defined in \EQn{eq:3vecnotCS2}: 

\begin{eqnarray}
D \dvv + \dvv  \left( \frac{1}{\tau_d} + \frac{D a_4}{2a_4} + \frac{\Theta}{2} \right) &=& \frac{1}{a_4}
\LB \Cv_k^{\rm eq} - \Cv_k \RB
\label{eq:IS_dvv} 
\end{eqnarray}
and
\begin{eqnarray}
D \evv + \evv \left( \frac{1}{\tau_e} + \frac{D a_5}{2a_5} + \frac{\Theta}{2} \right) &=& - \frac{\Cv_\omega}{2 a_5}.
\label{eq:IS_evv}
\end{eqnarray}
Since $1/(2 a_4) = \lambda T/\tau_d$ and $-1/(2 a_5) = \gamma T/\tau_e$, the right-hand side of the radial component of \EQ{eq:IS_dvv} equals $2 \lambda T (C^{\rm eq}_{kr} - C_{kr})/\tau_d$. Hence, in the limit $\tau_d \to 0$ and $\tau_e \to 0$, \EQSTWO{eq:IS_dvv}{eq:IS_evv} formally reproduce the Navier--Stokes expressions \EQSTWOn{eq:CS_NS_dr}{eq:CS_NS_e}.

\begin{figure}[t]
    \centering
    \includegraphics[width=0.5\columnwidth]{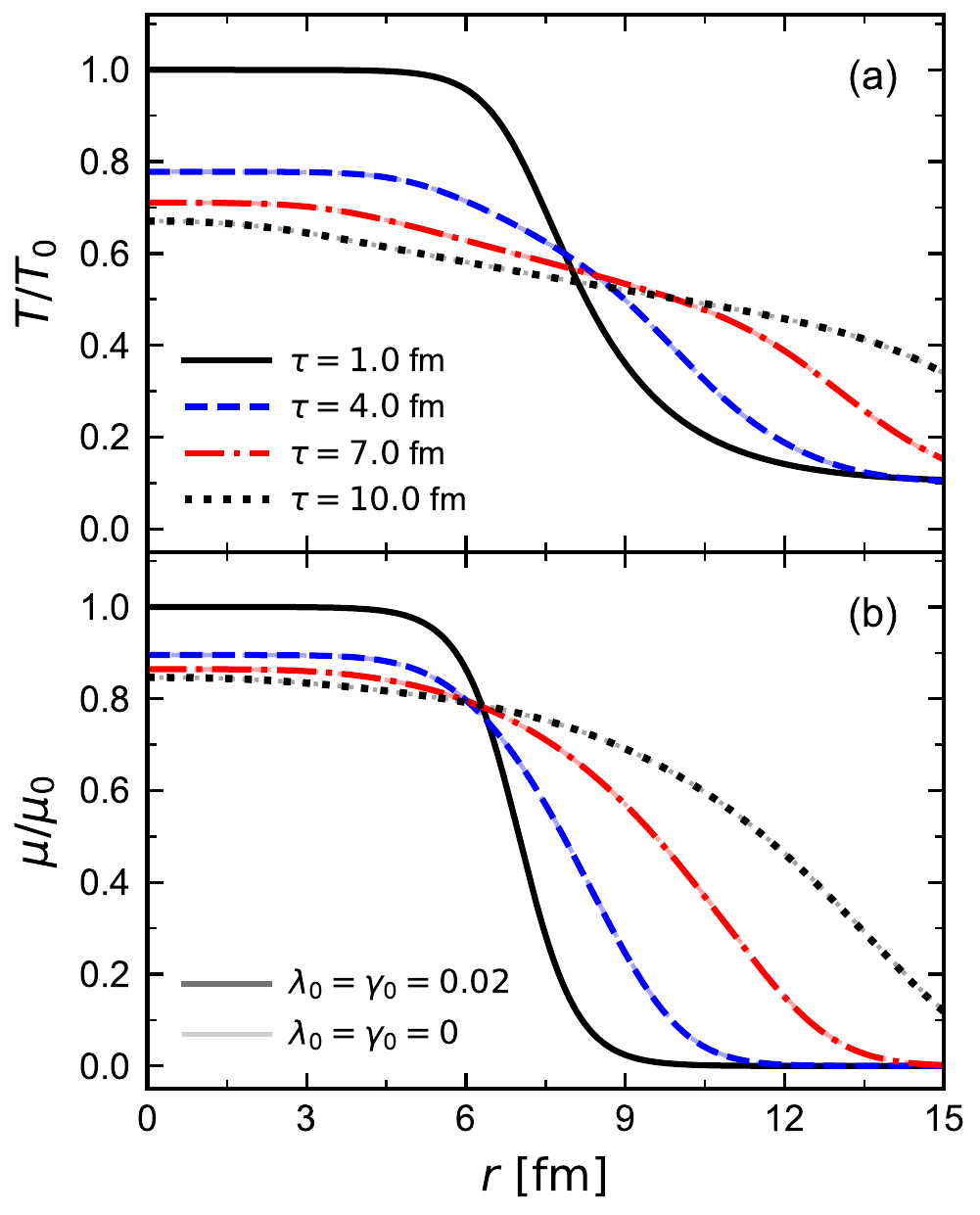}
    \caption{Radial profiles of (a) the temperature $T$ and (b) the baryon chemical potential $\mu$, normalized to their initial values at the center, $T_0$ and $\mu_0$. The initial conditions are given by \EQ{eq:CS_init}. Different line styles correspond to different values of the proper time $\tau$, as described in the legend. Opaque lines show the results of the dissipative evolution with $\lambda_0 = \gamma_0 = 0.02$ and $\tau_d = \tau_e = 1.0$~fm, while semitransparent lines show the evolution without dissipation, $\lambda_0 = \gamma_0 = 0$. The two types of lines overlap, and no difference between dissipative and perfect-fluid evolutions can be noticed in this case.}
    \label{fig:cs_T_mu}
\end{figure}

Before we switch to a discussion of our numerical results based on the IS method, let us come back to the issues discussed in the end of the last section. First of all, we notice that similar problems with overdetermined systems of equations occur in the IS approach, which forces us to assume that $C_{k\phi} = C_{\omega z} = d_\phi = e_z = 0$ and  $C_{kz} = C_{\omega \phi} = d_z = e_\phi = 0$. Thus, below we restrict ourselves to study the dynamics in the radial sector only. Second, we observe that the IS theory gives us the time derivative of $d_r$ that can be directly used in \EQSTWO{eq:E0CS}{eq:E1CS}. In this way, we naturally avoid problems of the NS method.


Let us turn now to a  discussion of our numerical results obtained for the CS model in the IS version. The initial proper time is $\tau_0 = 1.0$~fm, and the evolution is followed up to the final time $\tau_f = 10.0$~fm. As in \CIT{Drogosz:2026egr}, the initial transverse profiles follow the Woods--Saxon distribution $\rho(r) = \LSB 1 + \exp\LB (r - R_{\rm WS})/a_{\rm WS} \RB \RSB^{-1}$ with $R_{\rm WS} = 7.0$~fm and $a_{\rm WS} = 0.54$~fm, namely
\begin{equation}
T(\tau_0, r) = (T_0 - T_{\rm vac})\, \rho(r)^{1/3} + T_{\rm vac}, \qquad
\mu(\tau_0, r) = \mu_0\, \rho(r), \qquad
C_{\omega r}(\tau_0, r) = \rho(r),
\label{eq:CS_init}
\end{equation}    
with $T_0 = 0.3$~GeV, $\mu_0 = 0.03$~GeV, and $T_{\rm vac} = 0.03$~GeV. The fluid is initially at rest, $\theta(\tau_0, r) = 0$, and $C_{kr}(\tau_0, r) = d_r(\tau_0, r) = e_r(\tau_0, r) = 0$. As in the TH case, $m = 1.116$~GeV, $\lambda = \lambda_0 \sigma$ and $\gamma = \gamma_0 \sigma$ with $\lambda_0 = \gamma_0 = 0.02$, and $\tau_d = \tau_e = 1.0$~fm. We note that the electriclike sector ($C_{kr}$, $d_r$) is not coupled to the magneticlike one ($C_{\omega r}$, $e_r$); therefore, the results for $C_{kr}$ and $d_r$ depend neither on the initial value of $C_{\omega r}$ nor on the relaxation time $\tau_e$.

\begin{figure}[t]
    \centering
    \includegraphics[width=0.5\columnwidth]{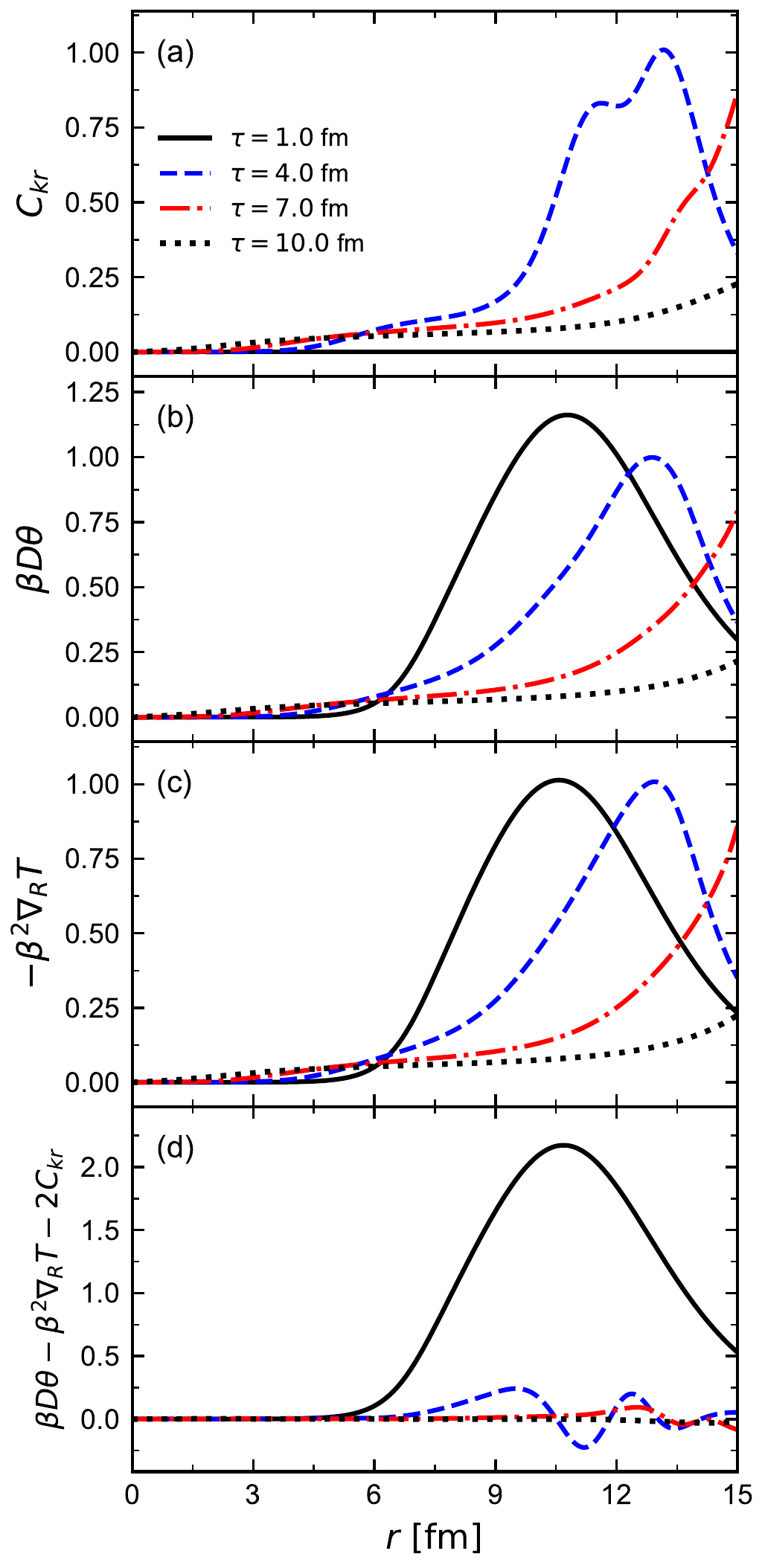}
    \caption{Radial profiles of (a) the radial electriclike spin polarization component $C_{kr}$, (b) the term $\beta D\theta$, (c) the term $-\beta^2 \nabla_R T$, and (d) the combination $\beta D\theta - \beta^2 \nabla_R T - 2C_{kr}$ that appears on the right-hand side of \EQ{eq:IS_dvv}. Different line styles correspond to different values of the proper time $\tau \in \{1.0~\text{fm}, 4.0~\text{fm}, 7.0~\text{fm}, 10.0~\text{fm}\}$. The system is initialized with $C_{kr}(\tau_0, r) = d_r(\tau_0, r) = 0$, $T_0 = 0.3$~GeV, and $\mu_0 = 0.03$~GeV; the kinetic coefficient is $\lambda = \lambda_0 \sigma$ with $\lambda_0 = 0.02$, and the relaxation time is $\tau_d = 1.0$~fm.}
    \label{fig:cs_Ckr_decomposition}
\end{figure}

\begin{figure}[t]
    \centering
    \includegraphics[width=0.5\textwidth]{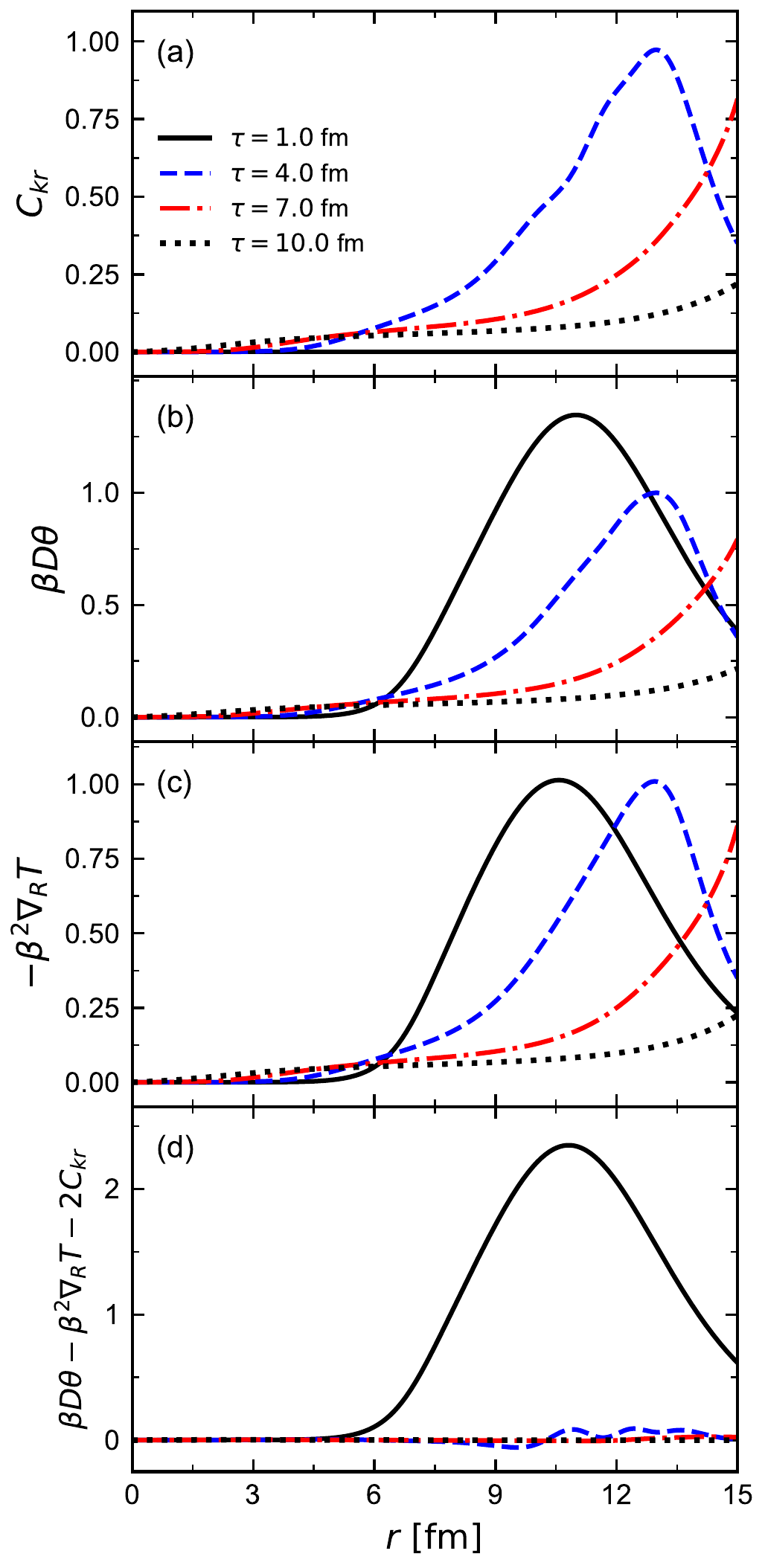}
    \caption{Same as \FIG{fig:cs_Ckr_decomposition}, except for the relaxation time $\tau_d = 0.5$~fm. }
    \label{fig:cs_Ckr_taud}
\end{figure}

The time evolution of the temperature and baryon chemical potential is shown in~\FIG{fig:cs_T_mu}. The opaque lines show the results with dissipation, and the semitransparent lines describe our previous results obtained for the perfect spin fluid~\CITn{Drogosz:2026egr}. The fact that the two types of lines overlap and are, in practice, indistinguishable means that the dissipative terms involving $d_r$ can be neglected in~\EQSTWOn{eq:E0CS}{eq:E1CS}. Hence, in the considered case, one can separate a spinless hydrodynamic background from the spin dynamics.~\footnote{This might be not true for other choices of parameters.}

Our numerical results for the spin electric radial sector are shown in~\FIG{fig:cs_Ckr_decomposition}. Panel (a) shows spatial profiles of the coefficient $C_{kr}$ for different evolution times. Panels (b) and (c) show the two contributions to the coefficient $C^{\rm eq}_{kr}$; see \EQ{eq:varpiCS}: the term $\beta D\theta$, connected with the acceleration of the fluid, and the term $-\beta^2 \nabla_R T$, connected with the radial temperature gradient. Both terms are positive and of a similar magnitude. They are localized in the region of the expansion front, where the fluid is accelerated and the temperature drops, and they move together with the front towards larger values of $r$. Panel (d) shows the spatial dependence of the combination \mbox{$\beta D\theta - \beta^2 \nabla_R T - 2 C_{kr} = 2 \LB C^{\rm eq}_{kr} - C_{kr} \RB$}.

At the initial time, $C_{kr} = 0$, and the right-hand side of \EQ{eq:IS_dvv}, shown in panel (d), is determined by thermal vorticity alone. This source generates the electric dissipation $d_r$, which in turn builds up the coefficient $C_{kr}$ through the spin equation~\EQn{eq:SaCS}. As a result, the profile of $C_{kr}$ in panel (a) quickly adopts the shape of $C^{\rm eq}_{kr}$, and the combination $\beta D\theta - \beta^2 \nabla_R T - 2 C_{kr} = 2 \LB C^{\rm eq}_{kr} - C_{kr} \RB$ becomes small compared to its initial value and continues to decrease during the evolution. Thus, $C_{kr}$ does not relax to a constant value but follows the time-dependent thermal vorticity. This behavior agrees with the form of the entropy production \EQn{eq:CS_divS}, which vanishes for $C_{kr} = C^{\rm eq}_{kr}$.

\begin{figure}[t]
    \centering
    \includegraphics[width=0.5\columnwidth]{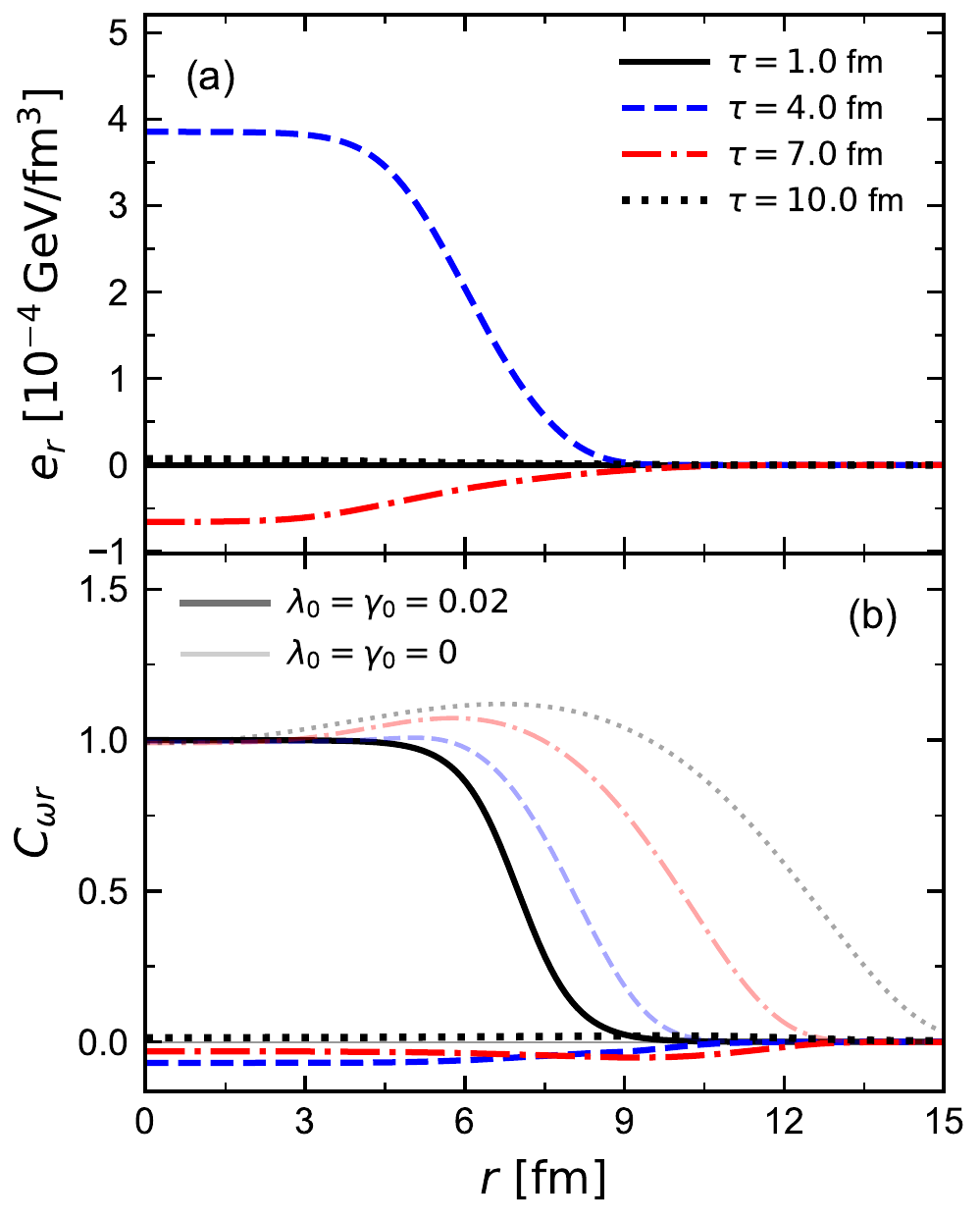}
    \caption{Radial profiles of (a) the radial component $e_r$ of the magnetic dissipation vector and (b) the radial magneticlike spin polarization coefficient $C_{\omega r}$. The initial conditions are given by \EQ{eq:CS_init} with $e_r(\tau_0, r) = 0$. Different line styles correspond to different values of the proper time $\tau$, as described in the legend. Opaque lines show the results of the dissipative evolution with $\gamma_0 = 0.02$ and $\tau_e = 1.0$~fm, while semitransparent lines in panel (b) show the evolution without dissipation, $\lambda_0 = \gamma_0 = 0$, for which $e_r = 0$.}
    \label{fig:cs_er_Cwr}
\end{figure}

Figure~\ref{fig:cs_Ckr_taud} shows the time dependence of the same quantities as in \FIG{fig:cs_Ckr_decomposition} but obtained here with a smaller relaxation time value, $\tau_d = 0.5$~fm. Once again, we observe the spin polarization tending toward the value corresponding to thermal vorticity; this process occurs more rapidly due to the smaller value of $\tau_d$ assumed. 

It is worth noting that the similar behavior of the functions $\beta D\theta$ and $-\beta^2 \nabla_R T$, shown in \FIG{fig:cs_Ckr_decomposition} and \FIG{fig:cs_Ckr_taud}, is not accidental. In the case of the perfect fluid, the hydrodynamic equations have the form $\p_\mu \LB T u^\nu \RB = \p^\nu T$, which implies an exact equality between these two functions. The case considered here includes a small chemical potential and spin effects, which leads to differences between $\beta D\theta$ and $-\beta^2 \nabla_R T$, although they are rather small. 

Finally, in~\FIG{fig:cs_er_Cwr} we show the time evolution of the coefficients $e_r$ and $C_{\omega r}$. As expected, both $e_r$ and $C_{\omega r}$ approach zero. For comparison, we added semitransparent lines showing the perfect spin-fluid results. If dissipation is not taken into account, the value of the coefficient $C_{\omega r}$ increases slowly with time. 

\section{Summary}
\label{sec:summary}

To summarize our findings, we make the following observations:

\begin{itemize}

\item[1.] Inclusion of the antisymmetric parts of the energy--momentum tensor implies that dissipative dynamics drives the spin polarization tensor $\omega$ toward thermal vorticity $\varpi$. The speed of this process depends on the values of the spin kinetic coefficients and spin relaxation times. 

\item[2.] At the same time, a transfer between the spin and orbital parts of the total angular momentum takes place. This process can be interpreted as a dissipative spin-orbit interaction. 

\item[3.] Antisymmetric parts of the energy--momentum tensor appear in the equations of motion (conservation laws for energy and momentum), which, in the cases under consideration, implies that certain components of the spin-polarization tensor must vanish. These constraints become increasingly stringent as the symmetries imposed on the hydrodynamic flow become more restrictive.

\item[4.] The presence of the time derivatives of the antisymmetric parts of the energy--momentum tensor in the equations of motion leads to problems with the application of the Navier--Stokes prescription. They are automatically solved if the Israel--Stewart method is applied. 

\item[5.] The experience gained from the analyses of boost-invariant systems considered in this work can be used in more realistic calculations based on spin hydrodynamics. 

\end{itemize}

\medskip
\begin{acknowledgments}

This work was supported in part by the National Science Centre, Poland (NCN) Grant No.~2022/47/B/ST2/01372. 
\end{acknowledgments}

\begin{appendix}

\section{Useful identities for boost-invariant and cylindrically symmetric systems}
\label{sec:BICSuseful}

\begin{align}
    \p_\mu U^\mu &= \f{\cosh\theta}{\tau} + \f{\sinh\theta}{r} + \sinh\theta\, \dot\theta + \cosh\theta\, \theta',\nn\\
    \p_\mu R^\mu &= \f{\sinh\theta}{\tau} + \f{\cosh\theta}{r} + \cosh\theta\, \dot\theta + \sinh\theta\, \theta',\nn\\
    \p_\mu Z^\mu &= 0,\nn\\
    \p_\mu \Phi^\mu &= 0, 
\end{align}
\begin{align}
    U^\mu \p_\mu f\LB\tau, \eta, r, \phi \RB &= \LB\cosh\theta \,\p_\tau + \sinh\theta\, \p_r\RB f\LB\tau, \eta, r, \phi\RB,\nn\\
    R^\mu \p_\mu f\LB\tau, \eta, r, \phi \RB &= \LB\sinh\theta\, \p_\tau + \cosh\theta\, \p_r\RB f\LB\tau, \eta, r, \phi\RB,\nn\\
    Z^\mu \p_\mu f(\tau, \eta, r, \phi) &= \f{1}{\tau}\, \p_\eta f(\tau, \eta, r, \phi),\nn\\
    \Phi^\mu \p_\mu f(\tau, \eta, r, \phi) &= \f{1}{r}\, \p_\phi f(\tau, \eta, r, \phi),
\end{align}
\begin{align}
    U^\mu \p_\mu U^\nu &= \LB\cosh\theta\, \dot\theta + \sinh\theta\, \theta'\RB R^\nu,\nn\\
    U^\mu \p_\mu R^\nu &=  \LB\cosh\theta\, \dot\theta + \sinh\theta\, \theta'\RB U^\nu,\nn\\
    U^\mu \p_\mu Z^\nu &= 0 ,\nn\\
    U^\mu \p_\mu \Phi^\nu &= 0,
\end{align}
\begin{align}
    R^\mu \p_\mu U^\nu &= \LB\sinh\theta\, \dot\theta + \cosh\theta\, \theta'\RB R^\nu,\nn\\
    R^\mu \p_\mu R^\nu &=  \LB\sinh\theta\, \dot\theta + \cosh\theta\, \theta'\RB U^\nu,\nn\\
    R^\mu \p_\mu Z^\nu &= 0,\nn\\
    R^\mu \p_\mu \Phi^\nu &= 0,
\end{align}
\begin{align}
    Z^\mu \p_\mu U^\nu &= \f{\cosh\theta}{\tau} Z^\nu,\nn\\
    Z^\mu \p_\mu R^\nu &= \f{\sinh\theta}{\tau} Z^\nu,\nn\\
    %
    %
    Z^\mu \p_\mu \Phi^\nu &= 0,
\end{align}
\begin{align}
    \Phi^\mu \p_\mu U^\nu &= \f{\sinh\theta}{r} \Phi^\nu,\nn\\
    \Phi^\mu \p_\mu R^\nu &= \f{\cosh\theta}{r} \Phi^\nu,\nn\\
    \Phi^\mu \p_\mu Z^\nu &= 0,  
    %
\end{align}
\begin{align}
    \na^\mu = \Delta^{\mu\nu} \p_\nu = -R^\mu\LB \sinh\theta\, \p_\tau + \cosh\theta\, \p_r \RB  - Z^\mu\f{1}{\tau}\, \p_\eta - \Phi^\mu \f{1}{r}\, \p_\phi,
\end{align}
\begin{align}
    \na^\mu T(\tau, r) = - \LB \sinh\theta\, \dot T +\cosh\theta\, T' \RB R^\mu,
\end{align}
\begin{align}
    a^\mu = U^\al \p_\al U^\mu = \LB \cosh\theta\, \dot \theta +\sinh\theta\, \theta' \RB R^\mu.
\end{align}

\end{appendix}



%

\end{document}